\documentclass[11pt]{article}
\usepackage[latin9]{inputenc}
\usepackage{textcomp}
\usepackage{amsmath}
\usepackage{amssymb}
\usepackage{graphicx}
\usepackage{esint}
\usepackage[unicode=true,
 bookmarks=false,
 breaklinks=false,pdfborder={0 0 1},backref=section,colorlinks=false]
 {hyperref}
\usepackage{breakurl}

\makeatletter

\DeclareFontEncoding{LGR}{}{}
\DeclareRobustCommand{\greektext}{%
  \fontencoding{LGR}\selectfont\def\encodingdefault{LGR}}
\DeclareRobustCommand{\textgreek}[1]{\leavevmode{\greektext #1}}
\ProvideTextCommand{\~}{LGR}[1]{\char126#1}

\providecommand{\tabularnewline}{\\}

\@ifundefined{date}{}{\date{}}

\usepackage{esint}
\usepackage{latexsym}
\usepackage{graphicx}\usepackage{bm}\usepackage{longtable}

\usepackage{xcolor}

\@addtoreset{equation}{section}

\makeatother

\begin{document}

\title{Holographic three flavor baryon in the Witten-Sakai-Sugimoto model
with the D0-D4 background}
\maketitle
\begin{center}
\footnote{Email: whcai@shu.edu.cn}Wenhe Cai\emph{$^{*}$} and \footnote{Email: siwenli@fudan.edu.cn}Si-wen
Li\emph{$^{\dagger}$}
\par\end{center}

\begin{center}
\emph{$^{\dagger}$Department of Physics,}\\
\emph{ Center for Field Theory and Particle Physics, }\\
\emph{ Fudan University, }\\
\emph{Shanghai 200433, China}\\
\par\end{center}

\begin{center}
\emph{$^{*}$Department of Physics, }\\
\emph{ Shanghai University, }\\
\emph{ Shanghai 200444, China}
\par\end{center}

\vspace{7mm}
\begin{abstract}
With the construction of the Witten-Sakai-Sugimoto model in the D0-D4
background, we systematically investigate the holographic baryon spectrum
in the case of three flavors. The background geometry in this model
is holographically dual to $U\left(N_{c}\right)$ Yang-Mills theory
in large $N_{c}$ limit involving an excited state with a nonzero
$\theta$ angle or glue condensate $\left\langle \mathrm{Tr}\mathcal{F}\wedge\mathcal{F}\right\rangle =8\pi^{2}N_{c}\tilde{\kappa}$,
which is proportional to the charge density of the smeared D0-branes
through a parameter $b$ or $\tilde{\kappa}$. The classical solution
of baryon in this model can be modified by embedding the Belavin-Polyakov-Schwarz-Tyupkin
(BPST) instanton and we carry out the quantization of the collective
modes with this solution. Then we extend the analysis to include the
heavy flavor and find that the heavy meson is always bound in the
form of the zero mode of the flavor instanton in strong coupling limit.
The mass spectrum of heavy-light baryons in the situation with single-
and double-heavy baryon is derived by solving the eigen equation of
the quantized collective Hamiltonian. Afterwards we obtain that the
constraint of stable baryon states has to be $1<b<3$ and the difference
in the baryon spectrum becomes smaller as the D0 charge increases.
It indicates that quarks or mesons can not form stable baryons if
the $\theta$ angle or glue condensate is sufficiently large. Our
work is an extension of the previous study of this model and also
agrees with those conclusions.
\end{abstract}
\newpage{}

\tableofcontents{}

\section{Introduction}

In QCD, it is well-known that the spontaneous breaking of chiral symmetry
dominates the light quark sector (u, d, s) while the heavy quark (c,
b, t) is characterized by the heavy-quark symmetry \cite{key-1}.
As measured by \cite{key-2,key-3}, the origin of these symmetries
relates to the chiral doubling in heavy-light mesons \cite{key-4,key-5,key-6,key-7}.
Recently, flurry of experiments report some new physics involving
heavy-light multiquark states \cite{key-8,key-9,key-10,key-11,key-12,key-13,key-14,key-15,key-16,key-17}
which might be a priori outside many classifications the of quark
model. Thus it would be urgent to formulate a non-perturbative model
of QCD that includes both chiral and heavy quark symmetry. On the
other hand, there have also been many researches about the spontaneous
parity violation in QCD with the running of the RHIC in recent years
\cite{key-18,key-19,key-20,key-21,key-22}. People usually use a nonzero
$\theta$ angle in the action to theoretically describe the $P$ or
$CP$ violation. Accordingly, metastable state with nonzero $\theta$
angle might probably be created in the hot and dense situation in
RHIC when the deconfinement transition happens in QCD. After a very
short time, these bubble forms with odd $P$ or $CP$ parity would
decay into the true vacuum soon \footnote{The recently holographic approach \cite{key-22+1} also supports this
statement.}. For comprehensive reviews, we refer the readers to \cite{key-23,key-24,key-25}
which is a proposal about the chiral magnetic effect (CME) as a test
of such phenomena about $P$ or $CP$ violation in hot QCD. In this
sense, the $\theta$ dependence of some observables in QCD or in the
gauge theory would be theoretically interesting at least. So several
present works have shown some properties of the $\theta$ dependence
in the gauge field theory e.g. $\theta$ dependence of deconfinement
transition \cite{key-26,key-27}, $\theta$ dependence in the spectrum
of the glueball \cite{key-28}, $\theta$ dependence in the large
$N$ limit \cite{key-29} and we strongly recommend \cite{key-30}
as an excellent review about the $\theta$ dependence in the gauge
field theory.

Since QCD in the low energy region becomes non-perturbative, the framework
based on the holographic construction by the gauge/gravity duality
provides an approach to investigate the aspects of the strongly coupled
gauge theory \cite{key-31,key-32}. Among various approaches to the
holographic duality of large $N_{c}$ QCD, a top-down model proposed
by Sakai and Sugimoto \cite{key-05,key-34} as an extension of Witten's
work \cite{key-04} (WSS model) is the most successful one since it
almost contains all necessary ingredients of QCD, e.g. baryons \cite{key-36,key-37},
chiral/deconfinement transitions \cite{key-38,key-39,key-40,key-41},
glueball spectrum and its interaction\cite{key-42,key-43,key-44,key-45,key-46}.
Specifically, flavors are introduced by embedding a stack of $N_{f}$
pairs of D8 and anti D8-branes ($\mathrm{D8}/\overline{\mathrm{D8}}$-branes)
as probes into the bubble geometry produced by $N_{c}$ D4-branes,
so this model describes the $N_{f}$ massless quarks by constructing
the configuration of $N_{c}$ D4- and $N_{f}$ $\mathrm{D8}/\overline{\mathrm{D8}}$-branes
in type IIA string theory. The $N_{f}$ chiral quark states in the
fundamental representation are identified as the massless spectrum
of the open strings stretched between the $N_{c}$ D4- and $N_{f}$
$\mathrm{D8}/\overline{\mathrm{D8}}$-branes. Particularly, there
is a naturally geometrical description of the spontaneous breaking
of the chiral symmetry in this model since the flavor branes are connected
at the bottom of the bubble as illustrated in Figure \ref{fig:1}.
So the separated $\mathrm{D}8/\overline{\mathrm{D}8}$-branes far
away combining near the bottom of the bubble can be interpreted as
the spontaneously breaking of $U_{R}\left(N_{f}\right)\times U_{L}\left(N_{f}\right)$
symmetry to $U_{V}\left(N_{f}\right)$ in the dual field theory. Additionally,
baryon in this model could be identified as a D4-brane warped on $S^{4}$,
namely \textquotedblleft baryon vertex\textquotedblright \footnote{To distinguish the $N_{c}$ D4-branes, we use ``$\mathrm{D4}^{\prime}$-brane''
to denote a baryon vertex from now on. }. And it can be effectively described by the instanton configuration
of the gauge field on the flavor branes \cite{key-09}. Then in order
to involve the topological $\theta$ term in the dual theory, it has
been recognized that the instantonic D-brane (D-instanton) relates
to the $\theta$ angle in holography by the construction of the string
theory \cite{key-48}. Hence the $\theta$ dependence could be introduced
to the WSS model in this way, that is adding the D-instanton (D0-branes)
to the $N_{c}$ D4-brane background geometry as in \cite{key-03,key-06}.
The systematical study of the WSS model in the D0-D4 brane background
(i.e. D0-D4/D8 brane system) can be reviewed in \cite{key-01,key-02}
and we can accordingly investigate the properties of the $\theta$
dependence in QCD or Yang-Mills theory holographically e.g.\cite{key-07,key-011,key-012,key-56}\footnote{There are also some other applications of the D0-D4 background in
holography e.g. \cite{key-013,key-014,key-59,key-60} where the setup
may be a little different from the model we introduced in this paper. }.

The purpose of this paper is to study the three-flavor baryon spectrum
with $\theta$ dependence or glue condensation in a holographic approach
i.e. using the WSS model in the D0-D4 brane background. Notice that
we have studied the $N_{f}=2$ two-flavor case with this model in
\cite{key-07,key-015}, so it would be natural to extend the present
analysis to the case of three flavors $N_{f}=3$. The main content
of this manuscript consists of two parts. In the first part, through
the analysis in \cite{key-028}, we search for a instanton solution
for the flavored gauge field in the situation with $N_{f}=3$ by including
the D0-branes. Then following \cite{key-07} and employing the soliton
picture, we derive the effective Hamiltonian for the collective modes
of baryon. After quantization, the baryon spectrum can be obtained
and all the calculations are done in the strong coupling limit. In
the second part, we extend our analysis to involve the heavy flavor
in the baryon spectrum. The heavy flavor could be introduced into
this model by embedding one pair of probe flavor branes (named as
\textquotedblleft heavy flavor brane\textquotedblright ) separated
from the other $N_{f}$ coincident flavor branes (named as \textquotedblleft light
flavor brane\textquotedblright ) with a heavy-light string (HL-string)
stretched between them \cite{key-015,key-016,key-017,key-018,key-019}
(as shown in Figure \ref{fig:2}). The low energy modes of HL-strings
could be identified as the heavy-light mesons and they could be approximated
in the bi- fundamental representation by the local vector fields in
the vicinity of the light flavor brane. Due to the finite separation
of the heavy and light flavor branes, the HL-string has nonzero vacuum
expectation value (vev) and the heavy-light fields therefore get mass
by the moduli span of the dilaton in the action. As a result, we can
evaluate the effective action involving the heavy flavor from the
dynamics of the light flavor brane since the baryon vertex lives inside
the light flavor brane. And this setup also develops the approach
of bound state in the context of the Skyrme model (e.g. \cite{key-67})
in holography.

The outline of this paper is as follows. In section 2, we review the
D0-D4 system and its dual theory. In section 3, we search for a instanton
solution for the gauge fields on the embedded flavor branes in the
case of three flavors. Afterwards the classical mass of the soliton
is evaluated with the instanton solution. In section 4, the collective
modes and their quantization are systematically investigated and the
baryon spectrum is then obtained by solving the eigen equation of
the collective Hamiltonian. In section 5, we start to consider the
heavy flavor additional to the light flavor baryon spectrum. This
section includes the effective action, quantization and single/double
baryon spectra with the heavy flavor. In the section 6, we briefly
use our baryon spectra to fit the experimental data and give some
discussion. 

\section{The D0-D4 background and the dual field theory}

In this section, we will briefly review the D0-D4 background and its
dual field theory by following \cite{key-01,key-02}. In Einstein
frame, the IIA supergravity solution of black coincident $N_{c}$
D4-branes with $N_{0}$ smeared D0-branes is given as,

\begin{equation}
ds^{2}=H_{4}^{-\frac{3}{8}}\left[-H_{0}^{\frac{7}{8}}f\left(U\right)d\tau^{2}+H_{0}^{\frac{1}{8}}\delta_{\mu\nu}dx^{\mu}dx^{\nu}\right]+H_{4}^{\frac{5}{8}}H_{0}^{\frac{1}{8}}\left[\frac{dU^{2}}{f\left(U\right)}+U^{2}d\Omega_{4}^{2}\right].
\end{equation}
Here $\tau$ represents the compactified direction on a cycle with
the period $\beta$. Respectively the dilaton, Ramond-Ramond 2- and
4-form are given as,

\begin{equation}
e^{-\mathbf{\Phi}}=g_{s}^{-1}\left(\frac{H_{4}}{H_{0}^{3}}\right)^{\frac{1}{4}},\ \ f_{2}=\frac{\left(2\pi l_{s}\right)^{7}g_{s}N_{0}}{\omega_{4}V_{4}}\frac{1}{U^{4}H_{0}^{2}}dU\wedge d\tau,\ \ f_{4}=\frac{\left(2\pi l_{s}\right)^{3}N_{c}g_{s}}{\omega_{4}}\epsilon_{4},\label{eq:2}
\end{equation}
where $g_{s}$ denotes the string coupling and,

\begin{align}
H_{4}=1+\frac{U_{Q4}^{3}}{U^{3}},\ \ H_{0}=1+\frac{U_{Q0}^{3}}{U^{3}},\ \ f\left(U\right)=1-\frac{U_{KK}^{3}}{U^{3}},\nonumber \\
U_{Q0}^{3}=\frac{1}{2}\left(-U_{KK}^{3}+\sqrt{U_{KK}^{6}+\left(\left(2\pi\right)^{5}l_{s}^{7}g_{s}\tilde{\kappa}N_{c}\right)^{2}}\right),\nonumber \\
U_{Q4}^{3}=\frac{1}{2}\left(-U_{KK}^{3}+\sqrt{U_{KK}^{6}+\left(\left(2\pi\right)^{5}l_{s}^{7}g_{s}N_{c}\right)^{2}}\right).
\end{align}
We use $d\Omega_{4}$, $\epsilon_{4}$ and $\omega_{4}=8\pi^{2}/3$
to represent the line element, the volume form and the volume of a
unit $S^{4}$. $V_{4}$ represents the volume of the D4-branes and
$U_{KK}$ is the horizon of the radius coordinate. $N_{c}$, $N_{0}$
denotes the numbers of D4- and D0-branes respectively and D0-branes
are smeared in the directions of $x^{0},x^{1},x^{2},x^{3}$ as shown
in Table \ref{tab:1}. So the number density of the D0-branes can
be defined as $N_{0}/V_{4}$. According to \cite{key-03}, we have
required that $N_{0}$ is order of $N_{c}$ in order to take account
of the full backreaction from the D0-branes. Therefore $\tilde{\kappa}$
would be order of $\mathcal{O}\left(1\right)$ in the large $N_{c}$
limit which is defined as $\tilde{\kappa}=N_{0}/\left(N_{c}V_{4}\right)$.

In the string frame, interchanging $x^{0},\tau$  and taking the near
horizon limit $\alpha^{\prime}\rightarrow0$ with fixed $U/\alpha^{\prime}$
and $U_{KK}/\alpha^{\prime}$, we could obtain the D0-D4 bubble geometry
which is,

\begin{align}
ds^{2}= & \left(\frac{U}{R}\right)^{3/2}\left[H_{0}^{1/2}\eta_{\mu\nu}dx^{\mu}dx^{\nu}+H_{0}^{-1/2}f\left(U\right)d\tau^{2}\right]\nonumber \\
 & +H_{0}^{1/2}\left(\frac{R}{U}\right)^{3/2}\left[\frac{dU^{2}}{f\left(U\right)}+U^{2}d\Omega_{4}^{2}\right],\label{eq:4}
\end{align}
and the dilaton becomes,

\begin{equation}
e^{\mathbf{\Phi}}=g_{s}\left(\frac{U}{R}\right)^{3/4}H_{0}^{3/4},\label{eq:5}
\end{equation}
where $\alpha^{\prime}=l_{s}^{2}$, $R^{3}=\pi g_{s}l_{s}^{3}N_{c}$
is the limit of $U_{Q4}^{3}$ and $l_{s}$ represents the length of
the string. The spacetime ends at $U=U_{KK}$ in the bubble geometry
(\ref{eq:4}) as shown in Figure \ref{fig:1}. The period $\beta$
of $\tau$ must satisfy the following relation in order to avoid the
conical singularity at $U_{KK}$, which is,

\begin{equation}
\beta=\frac{4\pi}{3}U_{KK}^{-1/2}R^{3/2}b^{1/2},\ \ b\equiv H_{0}\left(U_{KK}\right).\label{eq:6}
\end{equation}

In the low-energy effective description, the dual theory is a five-dimensional
$U\left(N_{c}\right)$ Yang-Mills theory which lives inside the worldvolume
of D4-brane. Since one direction of the D4-brane is compactified on
a cycle $\tau$, the four-dimensional Yang-Mills coupling could be
obtained as studied in \cite{key-04}, which is relating the D4-brane
tension and the five-dimensional Yang-Mills coupling constant $g_{5}$,
then analyzing the relation of the compactified five-dimensional theory
and the four dimensions on the $\tau$ direction. Thus the resultant
four-dimensional Yang-Mills coupling is,

\begin{equation}
g_{YM}^{2}=\frac{g_{5}^{2}}{\beta}=\frac{4\pi^{2}g_{s}l_{s}}{\beta},
\end{equation}
$b$ and $R^{3}$ can be accordingly evaluated as,

\begin{equation}
b=\frac{1}{2}\left[1+\left(1+C\beta^{2}\right)^{1/2}\right],\ \ C\equiv\left(2\pi l_{s}^{2}\right)^{6}\lambda^{2}\tilde{\kappa}/U_{KK}^{6},\ \ R^{3}=\frac{\beta\lambda l_{s}^{2}}{4\pi},\label{eq:8}
\end{equation}
where the 't Hooft coupling $\lambda$ is defined as $\lambda=g_{YM}^{2}N_{c}$.
Hence it is natural to define a mass scale as $M_{KK}=2\pi/\beta$
so that the Kaluza-Klein (KK) modes can be introduced. In order to
break the supersymmetry in the low-energy theory, the anti-periodic
condition has to be imposed on the fermions as in \cite{key-05},
thus the scalar and fermion become massive below the KK mass scale.
Consequently, the massless modes of the open strings on the D4-branes,
which is described by a pure Yang-Mills theory, dominate the dynamics
in dual field theory. According to (\ref{eq:6}) and (\ref{eq:8}),
we can obtain the following relations,

\begin{equation}
\beta=\frac{4\pi\lambda l_{s}^{2}}{9U_{KK}}b,\ \ \ \ M_{KK}=\frac{9U_{KK}}{2\lambda l_{s}^{2}b}.\label{eq:9}
\end{equation}
 Due to $b\geq1$ and $U_{KK}\geq2\lambda l_{s}^{2}M_{KK}/9$, $\beta$
can be solved by using (\ref{eq:8}) and (\ref{eq:9}),

\begin{equation}
\beta=\frac{4\pi\lambda l_{s}^{2}}{9U_{KK}}\frac{1}{1-\frac{\left(2\pi l_{s}^{2}\right)^{8}}{81U_{KK}^{8}}\lambda^{4}\tilde{\kappa}^{2}},\ \ b=\frac{1}{1-\frac{\left(2\pi l_{s}^{2}\right)^{8}}{81U_{KK}^{8}}\lambda^{4}\tilde{\kappa}^{2}}.
\end{equation}

In the presence of the smeared D0-branes, let us consider a probe
D4-brane whose effective action takes the following form,

\begin{equation}
S_{D_{4}}=-\mu_{4}\mathrm{Tr}\int d^{4}xd\tau e^{-\mathbf{\Phi}}\sqrt{-\det\left(G+2\pi\alpha^{\prime}\mathcal{F}\right)}+\mu_{4}\int C_{5}+\frac{1}{2}\left(2\pi\alpha^{\prime}\right)^{2}\mu_{4}\int C_{1}\wedge\mathcal{F}\wedge\mathcal{F},\label{eq:11}
\end{equation}
where $\mu_{4}=\left(2\pi\right)^{-4}l_{s}^{-5}$, and $G$ is the
induced metric on the world volume of the D4-brane. $\mathcal{F}$
represents the gauge field strength on the D4-brane. We have used
$C_{5},\ C_{1}$ to denote the Ramond-Ramond 5- and 1- form respectively
and their field strengths are given in (\ref{eq:2}). Notice that
the leading-order expansion of the first term in (\ref{eq:11}) with
respect to sufficiently small $\mathcal{F}$ forms the Yang-Mills
action. In the bubble D0-D4 solution, we have $C_{1}\sim\theta d\tau$
according to (\ref{eq:2}), thus D0-branes are indeed D-instantons
(as shown in Table \ref{tab:1}), so the last term in (\ref{eq:11})
could be integrated out as\footnote{The explicit derivation of the $\theta$ angle and glue condensate
in the D0-D4 background can be respectively reviewed in \cite{key-06}
and \cite{key-01}. One can find that $\theta$ angle and glue condensate
have a same origin in this holographic system although they are different
physical concepts.},

\begin{equation}
\int_{S_{\tau}^{1}}C_{1}\sim\theta\sim\tilde{\kappa},\ \ \int_{S_{\tau}^{1}\times\mathbb{R}^{4}}C_{1}\wedge\mathcal{F}\wedge\mathcal{F}\sim\theta\int_{\mathbb{R}^{4}}\mathcal{F}\wedge\mathcal{F}.\label{eq:12}
\end{equation}
A free parameter $\tilde{\kappa}$ (related to the $\theta$ angle
in the dual field theory) has been introduced into the Witten-Sakai-Sugimoto
model by this string theory background, hence this background is not
dual to the vacuum state of the gauge field theory. As studied in
\cite{key-01,key-02,key-06}, it implies some excited states in the
dual field theory with a constant homogeneous field strength background
may be described in the D0-D4 model, or equivalently, the dual field
theory behaves like $\theta$-dependent Yang-Mils (YM) theory. By
following \cite{key-01,key-02,key-07}, we can evaluate the expectation
value of $\mathrm{Tr}\mathcal{F}\wedge\mathcal{F}$ as $\left\langle \mathrm{Tr}\mathcal{F}\wedge\mathcal{F}\right\rangle =8\pi^{2}N_{c}\tilde{\kappa}$.
So the deformed relations to the QCD variables in the presence of
D0-branes are given as,
\begin{equation}
R^{3}=\frac{\lambda l_{s}^{2}}{2M_{KK}},\ \ g_{s}=\frac{\lambda}{2\pi M_{KK}N_{c}l_{s}},\ \ U_{KK}=\frac{2}{9}M_{KK}\lambda l_{s}^{2}b.\label{eq:13}
\end{equation}

\section{The effective action from D-brane}

\subsection{D-brane setup}

The chiral symmetry in the D0-D4 system is $U_{R}\left(N_{f}\right)\times U_{L}\left(N_{f}\right)$
which can be introduced by adding a stack of probe $N_{f}$ D8-anti-D8
($\mathrm{D}8/\overline{\mathrm{D}8}$) branes. Usually they are named
as flavor branes. The separated $\mathrm{D}8/\overline{\mathrm{D}8}$-branes
far away combining near the bottom $U=U_{KK}$ can be geometrically
interpreted as the spontaneously breaking of $U_{R}\left(N_{f}\right)\times U_{L}\left(N_{f}\right)$
symmetry to $U_{V}\left(N_{f}\right)$ in the dual field theory. This
can be verified by the appearance of massless Goldstones \cite{key-80}.
The brane configurations are illustrated in Table \ref{tab:1}. 

\begin{table}
\begin{centering}
\begin{tabular}{|c|c|c|c|c|c|c|c|c|c|c|}
\hline 
 & 0 & 1 & 2 & 3 & 4$\left(\tau\right)$ & 5$\left(U\right)$ & 6 & 7 & 8 & 9\tabularnewline
\hline 
\hline 
$N_{0}$ smeared D0-branes & = & = & = & = & - &  &  &  &  & \tabularnewline
\hline 
$N_{c}$ D4-branes & - & - & - & - & - &  &  &  &  & \tabularnewline
\hline 
$N_{f}$ flavor branes $\mathrm{D}8/\overline{\mathrm{D}8}$ & - & - & - & - &  & - & - & - & - & -\tabularnewline
\hline 
Baryon vertex $\mathrm{D}4^{\prime}$-brane & - &  &  &  &  &  & - & - & - & -\tabularnewline
\hline 
\end{tabular}
\par\end{centering}
\caption{\label{tab:1}The D-brane configurations: \textquotedblleft =\textquotedblright{}
denotes the smeared directions, \textquotedblleft -\textquotedblright{}
denotes the world volume directions.}
\end{table}

\begin{figure}
\begin{centering}
\includegraphics[scale=0.5]{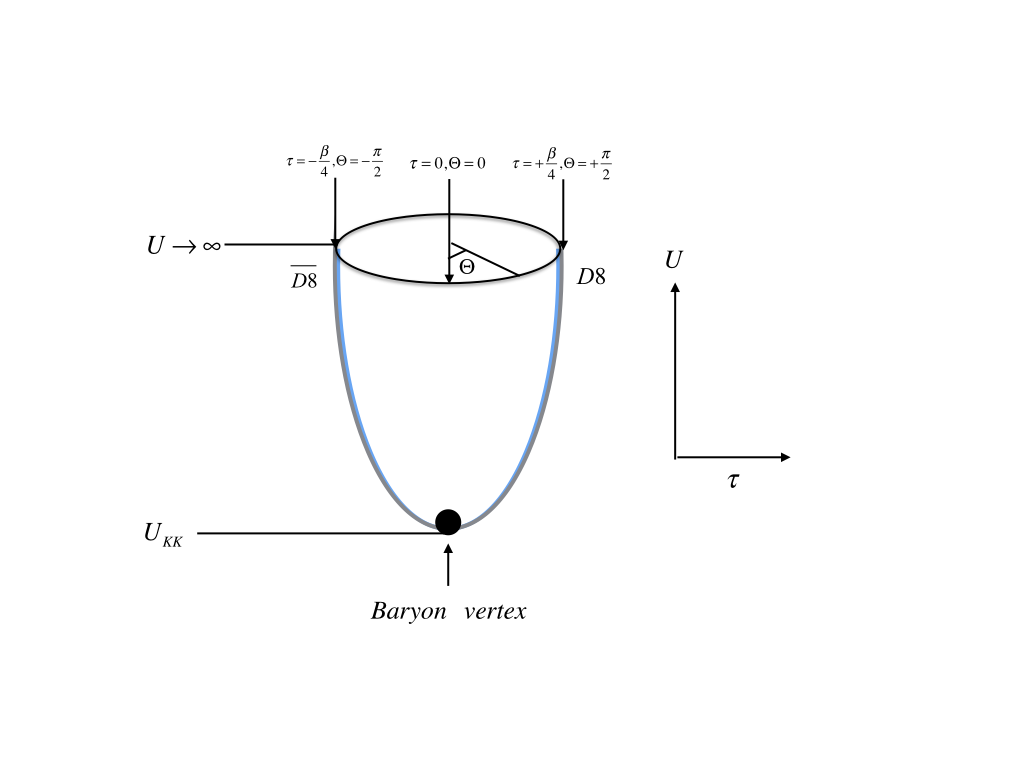}
\par\end{centering}
\caption{\label{fig:1}The bubble configuration of the D0-D4 geometry in the
$\tau-U$ plane where $\tau$ is compactified on $S^{1}$. This background
(cigar) is produced by $N_{c}$ D4-branes with $N_{0}$ smeared D0-branes.
The flavor $\mathrm{D}8/\overline{\mathrm{D}8}$-branes live at antipodal
position of the cigar represented by the blue line. Particularly we
will consider the case of $N_{f}=3$ in this paper. The baryon spectrum
is created by the baryon vertex ($\mathrm{D4}^{\prime}$-brane) living
inside the flavor branes.}

\end{figure}

Accordingly, the induced metric on the probe $\mathrm{D}8/\overline{\mathrm{D}8}$-branes
is,

\begin{align}
ds_{\mathrm{D}8/\overline{\mathrm{D}8}}^{2}= & \left(\frac{U}{R}\right)^{3/2}H_{0}^{-1/2}\left[f\left(U\right)+\left(\frac{R}{U}\right)^{3}\frac{H_{0}}{f\left(U\right)}U^{\prime2}\right]d\tau^{2}\nonumber \\
 & +\left(\frac{U}{R}\right)^{3/2}H_{0}^{1/2}\eta_{\mu\nu}dx^{\mu}dx^{\nu}+H_{0}^{1/2}\left(\frac{R}{U}\right)^{3/2}U^{2}d\Omega_{4}^{2}.
\end{align}
where $U^{\prime}$ denotes the derivative with respect to $\tau$.
The action of the $\mathrm{D}8/\overline{\mathrm{D}8}$-branes can
be obtained as,

\begin{equation}
S_{\mathrm{D}8/\overline{\mathrm{D}8}}\propto\int d^{4}xdUH_{0}\left(U\right)U^{4}\left[f\left(U\right)+\left(\frac{R}{U}\right)^{3}\frac{H_{0}}{f\left(U\right)}U^{\prime2}\right]^{1/2},\label{eq:15}
\end{equation}
then the equation of motion for $U\left(\tau\right)$ can be derived
as,

\begin{equation}
\frac{d}{d\tau}\left(\frac{H_{0}\left(U\right)U^{4}f\left(U\right)}{\left[f\left(U\right)+\left(\frac{R}{U}\right)^{3}\frac{H_{0}}{f\left(U\right)}U^{\prime2}\right]^{1/2}}\right)=0,\label{eq:16}
\end{equation}
which can be interpreted as the the conservation of the energy. With
the initial conditions $U\left(\tau=0\right)=U_{0}$ and $U^{\prime}\left(\tau=0\right)=0$,
the generic formula of the embedding function $\tau\left(U\right)$
can be solved as,

\begin{equation}
\tau\left(U\right)=E\left(U_{0}\right)\int_{U_{0}}^{U}dU\frac{H_{0}^{1/2}\left(U\right)\left(\frac{R}{U}\right)^{3/2}}{f\left(U\right)\left[H_{0}^{2}\left(U\right)U^{8}f\left(U\right)-E^{2}\left(U_{0}\right)\right]^{1/2}},\label{eq:17}
\end{equation}
where $E\left(U_{0}\right)=H_{0}\left(U_{0}\right)U_{0}^{4}f^{1/2}\left(U_{0}\right)$
and $U_{0}$ denotes the connected position of the $\mathrm{D}8/\overline{\mathrm{D}8}$-branes.
Following \cite{key-05,key-01}, we introduce the new coordinates
$\left(r,\Theta\right)$ and $\left(y,z\right)$ which satisfy,

\begin{align}
y=r\cos\Theta, & \ \ z=r\sin\Theta,\nonumber \\
U^{3}=U_{KK}^{3}+U_{KK}r^{2}, & \ \ \Theta=\frac{2\pi}{\beta}\tau=\frac{3}{2}\frac{U_{KK}^{1/2}}{R^{3/2}H_{0}^{1/2}\left(U_{KK}\right)}.
\end{align}
In this model, the probe $\mathrm{D}8/\overline{\mathrm{D}8}$-branes
are embedded at $\Theta=\pm\frac{1}{2}\pi$ respectively i.e. $y=0$
as illustrated in Figure \ref{fig:1}. Hence the embedding function
of the flavor branes is $\tau\left(U\right)=\frac{1}{4}\beta$ so
that we have $U^{3}=U_{KK}^{3}+U_{KK}z^{2}$ on the $\mathrm{D}8/\overline{\mathrm{D}8}$-branes\footnote{With the boundary condition as discussed in \cite{key-01,key-05},
$\tau\left(U\right)=\frac{1}{4}\beta$ is indeed a solution of (\ref{eq:16})
.}. Therefore the induced metric on the flavor branes becomes,

\begin{equation}
ds_{\mathrm{D}8/\overline{\mathrm{D}8}}^{2}=H_{0}^{1/2}\left(\frac{U}{R}\right)^{3/2}\eta_{\mu\nu}dx^{\text{\textmu}}dx^{\nu}+\frac{4}{9}\frac{U_{KK}}{U}\left(\frac{U}{R}\right)^{3/2}H_{0}^{1/2}dz^{2}+H_{0}^{1/2}\left(\frac{R}{U}\right)^{3/2}U^{2}d\Omega_{4}^{2}.\label{eq:19}
\end{equation}
With the approach presented in \cite{key-09,key-010}, the baryon
spectrum with two flavors in this system has been studied in \cite{key-07,key-011,key-012,key-013,key-014}.
Therefore it would be natural to extend the analysis into the three-flavor
case (i.e. $N_{f}=3$) in this paper.

\subsection{Yang-Mills and Chern-Simons action with generic flavors}

Since the baryon vertex lives inside the flavor branes, the concern
of this section is to study the effective dynamics of the baryons
on the $\mathrm{D}8/\overline{\mathrm{D}8}$-branes. So let us consider
the effective theory of $N_{f}$ probe $\mathrm{D}8/\overline{\mathrm{D}8}$-branes
in the background (\ref{eq:4}). The action of the flavor branes contains
two terms which are Yang-Mills (YM) and Chern-Simons (CS) action.
Respectively they are,

\begin{align}
S_{\mathrm{YM}}^{\mathrm{D}8/\overline{\mathrm{D}8}}= & -2\tilde{T}U_{KK}^{-1}\int d^{4}xdzH_{0}^{1/2}\mathrm{Tr}\left[\frac{1}{4}\frac{R^{3}}{U}\mathcal{F}_{\mu\nu}\mathcal{F}^{\mu\nu}+\frac{9}{8}\frac{U^{3}}{U_{KK}}\mathcal{F}_{\mu z}\mathcal{F}^{\mu z}\right],\nonumber \\
S_{\mathrm{CS}}^{\mathrm{D}8/\overline{\mathrm{D}8}}= & \frac{N_{c}}{24\pi^{2}}\mathrm{Tr}\int_{\mathbb{R}^{4+1}}\omega_{5}\left(\mathcal{A}\right)+S_{C_{7}}^{\mathrm{D}8/\overline{\mathrm{D}8}},\label{eq:20}
\end{align}
where\footnote{$T_{8}$ is the tension of the D8-branes.},

\begin{align}
\tilde{T}= & \frac{\left(2\pi\alpha^{\prime}\right)^{2}}{3g_{s}}T_{8}\omega_{4}U_{KK}^{3/2}R^{3/2}=\frac{M_{KK}^{2}\lambda N_{c}b^{3/2}}{486\pi^{3}},\nonumber \\
\omega_{5}\left(\mathcal{A}\right)= & \mathcal{A}\mathcal{F}^{2}-\frac{1}{2}\mathcal{A}^{3}\mathcal{F}+\frac{1}{10}\mathcal{A}^{5},\nonumber \\
S_{C_{7}}^{\mathrm{D}8/\overline{\mathrm{D}8}}= & 2\pi\alpha^{\prime}\mu_{8}\int_{\mathrm{D}8/\overline{\mathrm{D}8}}C_{7}\wedge\mathrm{Tr}\left[\mathcal{F}\right].
\end{align}
Here $\mathcal{A}$ is $U\left(N_{f}\right)$ gauge field and $\mathcal{F}$
is its field strength. We can decompose the $U\left(N_{f}\right)$
gauge field $\mathcal{A}$ and its gauge field $\mathcal{F}$ strength
into $U\left(1\right)$ part $\hat{A},\hat{F}$ and $SU\left(N_{f}\right)$
part $A,F$ as,

\begin{align}
\mathcal{A}= & \mathcal{A}_{\mu}dx^{\mu}+\mathcal{A}_{z}dz=A+\frac{1}{\sqrt{2N_{f}}}\hat{A}=A^{s}T_{s}+\frac{1}{\sqrt{2N_{f}}}\hat{A},\nonumber \\
\mathcal{F}= & d\mathcal{A}+i\mathcal{A}\wedge\mathcal{A}=F+\frac{1}{\sqrt{2N_{f}}}\hat{F},\label{eq:22}
\end{align}
where the index $\mu$ runs from 0 - 3 and $T_{s}$ are hermitian
generators of $SU\left(N_{f}\right)$ with the index $s=1,2...N_{f}^{2}-1$.
In order to simplify the calculations, we can further define the dimensionless
variables by the following replacement,

\begin{equation}
\left(x^{\mu},z\right)\rightarrow\left(x^{\mu}/M_{KK},zU_{KK}\right),\ \left(\mathcal{A}_{\mu},\mathcal{A}_{z}\right)\rightarrow\left(\mathcal{A}_{\mu}U_{KK},\mathcal{A}_{z}/M_{KK}\right).\label{eq:23}
\end{equation}
On the other hand, in the strongly coupled limit, we also need to
obtain the explicit expression with $\lambda$ by the rescaling in
\cite{key-09,key-07,key-028} which is,

\begin{align}
\left(x^{0},x^{M}\right) & \rightarrow\left(x^{0},\lambda^{1/2}x^{M}\right),\nonumber \\
\left(\mathcal{A}_{0},\mathcal{A}_{M}\right) & \rightarrow\left(\mathcal{A}_{0},\lambda^{-1/2}\mathcal{A}_{M}\right),\nonumber \\
\left(\mathcal{F}_{MN},\mathcal{F}_{0M}\right) & \rightarrow\left(\lambda^{-1}\mathcal{F}_{MN},\lambda^{-1/2}\mathcal{F}_{0M}\right),\label{eq:24}
\end{align}
where the index $M,N$ runs over $1,2,3,z$ and the $\mathcal{O}\left(\lambda^{-1}\right)$
terms would be neglected as in \cite{key-09,key-07,key-028}. So when
(\ref{eq:22}) (\ref{eq:23}) and (\ref{eq:24}) is imposed, the action
$S_{\mathrm{YM}}^{\mathrm{D}8/\overline{\mathrm{D}8}}$ can be expanded
as,

\begin{align}
S_{\mathrm{YM}}^{\mathrm{D}8/\overline{\mathrm{D}8}}= & -aN_{c}b^{3/2}\int d^{4}xdz\bigg[\frac{\lambda}{4}\left(F_{MN}^{a}\right)^{2}-\frac{bz^{2}}{2}\left(\frac{5}{12}-\frac{1}{4b}\right)\left(F_{ij}^{a}\right)^{2}+\frac{bz^{2}}{4}\left(1+\frac{1}{b}\right)\left(F_{iz}^{a}\right)^{2}-\left(F_{0M}^{a}\right)^{2}\nonumber \\
 & +\frac{\lambda}{4}\hat{F}_{MN}^{2}-\frac{bz^{2}}{2}\left(\frac{5}{12}-\frac{1}{4b}\right)\hat{F}_{ij}^{2}+\frac{bz^{2}}{4}\left(1+\frac{1}{b}\right)\hat{F}_{iz}^{2}-\frac{1}{2}\hat{F}_{0M}^{2}\bigg]+\mathcal{O}\left(\lambda^{-1}\right),\label{eq:25}
\end{align}
with the index $i,j=1,2,3$. 

Notice that the CS term $S_{C_{7}}^{\mathrm{D}8/\overline{\mathrm{D}8}}$
in (\ref{eq:20}) becomes $\mathcal{O}\left(\lambda^{-1}\right)$
in the strongly coupled limit. Thus by using (\ref{eq:22}) (\ref{eq:23})
and (\ref{eq:24}), we can obtain the explicit formula of the CS term
(\ref{eq:20}) which is,

\begin{align}
S_{\mathrm{CS}}^{\mathrm{D}8/\overline{\mathrm{D}8}}= & \frac{N_{c}}{24\pi^{2}}\int\omega_{5}^{SU\left(N_{f}\right)}\left(A\right)+\frac{N_{c}}{24\pi^{2}}\sqrt{\frac{2}{N_{f}}}\epsilon_{MNPQ}\int d^{4}xdz\bigg[\frac{3}{8}\hat{A}_{0}\mathrm{Tr}\left(F_{MN}F_{PQ}\right)\nonumber \\
 & -\frac{3}{2}\hat{A}_{M}\mathrm{Tr}\left(\partial_{0}A_{N}F_{PQ}\right)+\frac{3}{4}\hat{F}_{MN}\mathrm{Tr}\left(A_{0}F_{PQ}\right)+\frac{1}{16}\hat{A}_{0}\hat{F}_{MN}\hat{F}_{PQ}\nonumber \\
 & -\frac{1}{4}\hat{A}_{M}\hat{F}_{MN}\hat{F}_{PQ}+\mathrm{total\ derivatives}\bigg]+\mathcal{O}\left(\lambda^{-1}\right),\label{eq:26}
\end{align}
with the index $M,N,P,Q=1,2,3,z$ and $\epsilon_{123z}=1$.

\subsection{Classical soliton solution representing a baryon}

With the action (\ref{eq:25}) (\ref{eq:26}), we can obtain the equations
of motion (EOM) up to $\mathcal{O}\left(\lambda^{-1}\right)$ for
the gauge fields as,

\begin{align}
D_{M}F_{0M}+\frac{1}{64\pi^{2}ab^{3/2}}\sqrt{\frac{2}{N_{f}}}\epsilon_{MNPQ}\hat{F}_{MN}F_{PQ}\nonumber \\
+\frac{1}{64\pi^{2}ab^{3/2}}\epsilon_{MNPQ}\left\{ F_{MN}F_{PQ}-\frac{1}{N_{f}}\mathrm{Tr}\left(F_{MN}F_{PQ}\right)\right\} +\mathcal{O}\left(\lambda^{-1}\right) & =0,\nonumber \\
D_{N}F_{MN}+\mathcal{O}\left(\lambda^{-1}\right) & =0,\nonumber \\
\partial_{M}\hat{F}_{0M}+\frac{1}{64\pi^{2}ab^{3/2}}\sqrt{\frac{2}{N_{f}}}\epsilon_{MNPQ}\left\{ \mathrm{Tr}\left(F_{MN}F_{PQ}\right)+\frac{1}{2}\hat{F}_{MN}\hat{F}_{PQ}\right\} +\mathcal{O}\left(\lambda^{-1}\right) & =0,\nonumber \\
\partial_{N}\hat{F}_{MN}+\mathcal{O}\left(\lambda^{-1}\right) & =0.\label{eq:27}
\end{align}
The first two equations in (\ref{eq:27}) are the EOMs for $SU\left(N_{f}\right)$
part while the last two equations are EOMs for the $U\left(1\right)$
part. Then let us find a static soliton solution of the EOMs in (\ref{eq:27})
which could be able to represent a baryon. Since the baryon can be
identified as a $\mathrm{D}4^{\prime}$-brane wrapped on $S^{4}$
which is equivalent to instanton configuration on the $\mathrm{D}8/\overline{\mathrm{D}8}$-branes,
the solution of leading part $D_{N}F_{MN}=0$ carrying a unit baryon
number could be obtained by the embedding of $SU\left(2\right)$ Belavin-Polyakov-Schwarz-Tyupkin
(BPST) instanton solution in the flat space to $SU\left(N_{f}\right)$: 

\begin{equation}
A_{M}^{\mathrm{cl}}\left(x\right)=-if\left(x\right)g\left(x\right)\partial_{M}g\left(x\right)^{-1},\label{eq:28}
\end{equation}
where the function $f\left(x\right)$ and $g(x)$ are given as,

\begin{align}
f\left(x\right)=\frac{x^{2}}{x^{2}+\rho^{2}}, & \ \ x=\sqrt{\left(x^{M}-X^{M}\right)^{2}},\nonumber \\
g\left(x\right)=\left(\begin{array}{cc}
g^{SU\left(2\right)}\left(x\right) & 0\\
0 & \boldsymbol{1}_{N_{f}-2}
\end{array}\right), & \ \ g^{SU\left(2\right)}\left(x\right)=\frac{1}{x}\left[\left(z-Z\right)\boldsymbol{1}_{2}+i\left(x^{i}-X^{i}\right)\sigma_{i}\right].
\end{align}
Here we have used $\boldsymbol{1}_{N_{f}}$, $\sigma_{i}$ ($i=1,2,3$)
to represent the $N_{f}\times N_{f}$ identity matrix and three Pauli
matrices respectively. The constants $X^{M}=\left\{ X^{i},Z\right\} $
denotes the position of the instanton and $\rho$ represents its size.
Notice that these constants have also been rescaled as in (\ref{eq:23})
(\ref{eq:24}). The field strengths of (\ref{eq:28}) are given as,

\begin{equation}
F_{ij}^{\mathrm{cl}}=\frac{4\rho^{2}}{\left(x^{2}+\rho^{2}\right)^{2}}\epsilon_{ijk}t_{k},\ \ F_{iz}^{\mathrm{cl}}=\frac{4\rho^{2}}{\left(x^{2}+\rho^{2}\right)^{2}}t_{i},\label{eq:30}
\end{equation}
where $t_{i}=\left(\begin{array}{cc}
\sigma_{i} & 0\\
0 & 0
\end{array}\right)$ is the $SU\left(N_{f}\right)$ embedding of $\sigma_{i}$. Then the
EOM of the $U\left(1\right)$ field $\hat{A}_{M}$ gives the solution
$\hat{A}_{M}=0$ up to a gauge transformation. So the EOM for $\hat{A}_{0}$
becomes,

\begin{equation}
\partial_{M}^{2}\hat{A}_{0}+\sqrt{\frac{2}{N_{f}}}\frac{3\rho^{4}}{ab^{3/2}\pi^{2}\left(x^{2}+\rho^{2}\right)^{4}}=0,
\end{equation}
then we have its solution as,

\begin{equation}
\hat{A}_{0}^{\mathrm{cl}}=\sqrt{\frac{2}{N_{f}}}\frac{1}{8ab^{3/2}\pi^{2}}\frac{1}{x^{2}}\left[1-\frac{\rho^{4}}{\left(x^{2}+\rho^{2}\right)^{2}}\right].\label{eq:32}
\end{equation}
The present solution $\hat{A}_{0}^{\mathrm{cl}}$ is regular at $x=0$
and vanished at the infinity $x\rightarrow\infty$. Finally, let us
solve the EOM for $A_{0}$ to find a solution. For a generic $N_{f}$,
plugging (\ref{eq:28}) and (\ref{eq:32}) into the first equation
in (\ref{eq:27}), we have,

\begin{equation}
D_{M}^{2}A_{0}+\frac{3}{2ab^{3/2}\pi^{2}}\frac{\rho^{4}}{\left(x^{2}+\rho^{2}\right)^{2}}\left(\begin{array}{cc}
\boldsymbol{1}_{2}-\frac{2}{N_{f}}\boldsymbol{1}_{2} & 0\\
0 & -\frac{2}{N_{f}}
\end{array}\right)=0.
\end{equation}
Then $A_{0}$ can be accordingly solved as,

\begin{equation}
A_{0}^{\mathrm{cl}}=\frac{1}{16ab^{3/2}\pi^{2}}\frac{1}{x^{2}}\left[1-\frac{\rho^{4}}{\left(x^{2}+\rho^{2}\right)^{2}}\right]\left(\begin{array}{cc}
\boldsymbol{1}_{2}-\frac{2}{N_{f}}\boldsymbol{1}_{2} & 0\\
0 & -\frac{2}{N_{f}}
\end{array}\right).\label{eq:34}
\end{equation}
So the mass $M$ of the static soliton solution can be evaluated by
the relation $S_{\mathrm{onshell}}=-\int dtM$ with the solution (\ref{eq:28})
(\ref{eq:32}) and (\ref{eq:34}). It gives,

\begin{align}
M= & aN_{c}b^{3/2}\int d^{4}xdz\bigg[\frac{1}{4}\left(F_{MN}^{a}\right)^{2}-\frac{bz^{2}}{2}\left(\frac{5}{12}-\frac{1}{4b}\right)\left(F_{ij}^{a}\right)^{2}\lambda^{-1}\nonumber \\
 & +\frac{bz^{2}}{4}\left(1+\frac{1}{b}\right)\left(F_{iz}^{a}\right)^{2}\lambda^{-1}-\left(F_{0M}^{a}\right)^{2}\lambda^{-1}-\frac{\lambda^{-1}}{2}\hat{F}_{0M}^{2}\bigg]\nonumber \\
 & -\frac{N_{c}}{24\pi^{2}}\epsilon_{MNPQ}\int d^{4}xdz\left[\sqrt{\frac{2}{N_{f}}}\frac{3}{8}\hat{A}_{0}\mathrm{Tr}\left(F_{MN}F_{PQ}\right)+\frac{3}{4}\mathrm{Tr}\left(A_{0}F_{MN}F_{PQ}\right)\right]+\mathcal{O}\left(\lambda^{-1}\right)\nonumber \\
= & 8ab^{3/2}\pi^{2}N_{c}\lambda+8ab^{3/2}\pi^{2}N_{c}\left[\frac{3-b}{12}\left(2Z^{2}+\rho^{2}\right)+\frac{1}{320ab^{3}\pi^{4}\rho^{2}}\right]+\mathcal{O}\left(\lambda^{-1}\right).\label{eq:35}
\end{align}
For the stable solution, we have to minimize $M$ in order to determine
the values of $\rho$ and $Z$. They are,

\begin{equation}
\rho_{min}^{2}=\frac{\sqrt{15}}{20}\frac{1}{\sqrt{3-b}}\frac{1}{ab^{3/2}\pi^{2}},\ \ Z_{min}=0.\label{eq:36}
\end{equation}
So the mass formula (\ref{eq:35}) and size of the instanton (\ref{eq:36})
are all independent of $N_{f}$ while they depends on $b$ which relates
to the number density of D0-branes. We can further rewrite the expression
in terms of the original parameters which means $\rho$ has to be
rescaled as $\rho\rightarrow\lambda^{1/2}\rho$, hence the minimum
value of the mass of the soliton is given by,

\begin{equation}
M_{min}=8ab^{3/2}\pi^{2}N_{c}+\sqrt{\frac{3-b}{15}}N_{c}.
\end{equation}
Note that the size $\rho$ of the soliton becomes parametrically small
for large $\lambda$ with respect to the scale of the curvature. Hence
it implies the soliton energy density is concentrated in a small region
of the space which motivates our approach of the BPST instanton solution.
Although this is an approximation based on the assumption that the
CS term and the curvature (which are crucial in determining the size
$\rho$ of the soliton) do not alter the soliton field significantly,
the numerical computation in \cite{key-r1} strongly supports its
validity. Therefore the self-dual instanton configuration would be
a good approximation to describe the holographic baryons in the limitation
of the small-instanton size. Additionally it is also worth noticing
that the contributions of higher order derivative terms in the DBI
action, which has been dropped off in (\ref{eq:20}), might become
important in the limitation of the small-instanton size as mentioned
in \cite{key-09}. We leave this issue for future study and continue
analysis based on the YM action (\ref{eq:20}) in the rest of this
paper. 

\section{Dynamics of the collective modes }

\subsection{Collectivization}

In this section, let us briefly outline the collective mode of baryon
and its effective dynamics. In order to derive the Lagrangian of the
collective modes, we require the moduli of the solution to be time-dependent,
i.e.

\begin{equation}
X^{\alpha},a^{s}\rightarrow X^{\alpha}\left(t\right),a^{s}\left(t\right),\ \ s=1,2...8
\end{equation}
with $X^{\alpha}=\left\{ X^{M},\rho\right\} $ and $a^{s}$ refers
to the $SU\left(3\right)$ orientation. So the $SU\left(3\right)$
gauge field is assumed to transform as,

\begin{equation}
A\left(t,x\right)\rightarrow V\left(t,x\right)\left(A^{\mathrm{cl}}\left(t,x\right)-id\right)V\left(t,x\right)^{-1},
\end{equation}
where $A^{\mathrm{cl}}$ refers to the classical solution (\ref{eq:28})
(\ref{eq:34}) with time-dependent $X^{\alpha}$ and $V\left(t\right)\in SU\left(3\right)$.
Using (\ref{eq:32}), the $U\left(1\right)$ gauge field with time-dependent
$X^{\alpha}$ are given as,

\begin{equation}
\hat{A}_{M}\left(t,x\right)=0,\ \ \hat{A}_{0}\left(t,x\right)=\hat{A}_{0}^{\mathrm{cl}}\left(x,X^{\alpha}\left(t\right)\right).
\end{equation}
Accordingly, the gauge field strength can be obtained as,

\begin{align}
F_{MN} & =V\left(t,x\right)F_{MN}^{\mathrm{cl}}V\left(t,x\right)^{-1},\nonumber \\
F_{0M} & =V\left(t,x\right)\left(\dot{X}^{\alpha}\frac{\partial}{\partial X^{\alpha}}A_{M}^{\mathrm{cl}}-D_{M}^{\mathrm{cl}}\Phi-D_{M}^{\mathrm{cl}}A_{0}^{\mathrm{cl}}\right)V\left(t,x\right)^{-1},\label{eq:41}
\end{align}
where ``$D$'' denotes the covariant derivative operator, $\Phi$
is defined as $\Phi=-iV^{-1}\partial_{0}V$ and `` $\dot{}$ ''
represents the derivative respected to time. Then the first equation
in (\ref{eq:27}) implies,

\begin{equation}
D_{M}^{\mathrm{cl}}\left(\dot{X}^{N}\frac{\partial}{\partial X^{N}}A_{M}^{\mathrm{cl}}+\dot{\rho}\frac{\partial}{\partial\rho}A_{M}^{\mathrm{cl}}-D_{M}^{\mathrm{cl}}\Phi\right)=0.\label{eq:42}
\end{equation}
For the case of $N_{f}=3$, the solution for $\Phi$ can be written
as,

\begin{equation}
\Phi\left(t,x\right)=-\dot{X}^{N}\left(t\right)A_{N}^{\mathrm{cl}}\left(x,X^{\alpha}\left(t\right)\right)+\chi^{s}\left(t\right)\Phi_{s}\left(x,X^{\alpha}\left(t\right)\right),
\end{equation}
where the explicit form of $\Phi_{s}\left(x,X^{\alpha}\left(t\right)\right)$
can be found in \cite{key-028}. With the boundary condition $\Phi_{I}\left(x,X^{\alpha}\left(t\right)\right)\rightarrow T_{I}$
and $A_{M}^{\mathrm{cl}}\left(t,x\right)\rightarrow0$ as $z\rightarrow+\infty$,
we can further obtain,

\begin{equation}
\chi^{s}\left(t\right)=-2i\mathrm{Tr}\left[T_{s}\mathbf{a}\left(t\right)^{-1}\cdot\dot{\mathbf{a}}\left(t\right)\right],
\end{equation}
where $\mathbf{a}\left(t\right)\equiv a^{s}\left(t\right)T_{s}$.
Finally we find that $F_{0M}$ is given by,

\begin{equation}
F_{0M}=V\left(t,x\right)\left(\dot{X}^{N}F_{MN}^{\mathrm{cl}}+\dot{\rho}\frac{\partial}{\partial\rho}A_{M}^{\mathrm{cl}}-\chi^{s}D_{M}^{\mathrm{cl}}\Phi_{s}-D_{M}^{\mathrm{cl}}A_{0}^{\mathrm{cl}}\right)V\left(t,x\right)^{-1}.\label{eq:45}
\end{equation}
And the whole gauge field can be therefore expressed as,

\begin{equation}
\mathcal{A}\left(t,x\right)=V\left(t,x\right)\left[\mathcal{A}^{\mathrm{cl}}\left(x,X^{\alpha}\left(t\right)\right)+\Phi\left(t,x\right)dt\right]V\left(t,x\right)^{-1}.
\end{equation}

\subsection{Quantization}

The effective potential $U\left(X^{\alpha}\right)$ of the collective
modes can be obtained by the relation $S_{\mathrm{YM}}^{\mathrm{D}8/\overline{\mathrm{D}8}}+S_{\mathrm{CS}}^{\mathrm{D}8/\overline{\mathrm{D}8}}=-\int dtU\left(X^{\alpha}\right)$.
Inserting (\ref{eq:41}) (\ref{eq:45}) into action (\ref{eq:25})
(\ref{eq:26}), we have,

\begin{equation}
U\left(X^{\alpha}\right)=M-ab^{3/2}N_{c}\int d^{3}xdz\mathrm{Tr}\left(\dot{X}^{N}F_{MN}^{\mathrm{cl}}+\dot{\rho}\frac{\partial}{\partial\rho}A_{M}^{\mathrm{cl}}-\chi^{I}D_{M}^{\mathrm{cl}}\Phi_{I}\right)^{2}-L_{\mathrm{CS}}+\mathcal{O}\left(\lambda^{-1}\right),
\end{equation}
where the $L_{\mathrm{CS}}$ term is defined as,

\begin{equation}
S_{\mathrm{CS}}^{\mathrm{D}8/\overline{\mathrm{D}8}}\left[\mathcal{A}\right]-S_{\mathrm{CS}}^{\mathrm{D}8/\overline{\mathrm{D}8}}\left[\mathcal{A}^{\mathrm{cl}}\right]=\int dtL_{\mathrm{CS}}.\label{eq:48}
\end{equation}
However, if we use the Chern-Simons term (\ref{eq:26}) for the case
of $N_{f}>2$, it fails to give the exact transformation law under
the gauge or chiral transformations. Particularly, (\ref{eq:26})
is unable to reproduce the important constraint of the hypercharge,

\begin{equation}
J_{8}=\frac{N_{c}}{2\sqrt{3}}.\label{eq:49}
\end{equation}
This issue has been revisited in \cite{key-029} where the contribution
at the boundary was added to (\ref{eq:26}). Resultantly, a new Chern-Simons
term was proposed in \cite{key-029} whose formula is\footnote{Since only the first term in (\ref{eq:20}) survives in the large
$\lambda$ limit, we consider the boundary contributions based on
this term in the following calculation.},

\begin{equation}
S_{\mathrm{CS}}^{\mathrm{new}}=S_{\mathrm{CS}}+\int_{N_{5}}\frac{1}{10}\mathrm{Tr}\left(h^{-1}dh\right)^{5}+\int_{\partial M_{5}}\alpha_{4}\left(dh,A\right),\label{eq:50}
\end{equation}
where the gauged 4-form $\alpha_{4}$ is given in \cite{key-029}
and $S_{\mathrm{CS}}$ refers to the CS term (\ref{eq:20}). Here
$N_{5}$ denotes a 5-dimensional manifold whose boundaries are defined
as $\partial N_{5}=\partial M_{5}=M_{4+\infty}-M_{4-\infty}$ with
the the asymptotic gauge field on the flavor branes,

\begin{equation}
A\big|_{z\rightarrow\pm\infty}=\bar{A}^{\pm}=h^{\pm}\left(d+A\right)h^{\pm-1},
\end{equation}
where $h\big|_{\partial M_{5}}=\left(h^{+},h^{-}\right)$ and $\bar{A}^{\pm}$
denotes the external gauge field. Notice that $A$ is required to
be well defined gauge field throughout $M_{5}$ and produces no-boundary
contributions. So it implies that the information about the topology
is kept at the holographic boundary $z\rightarrow\pm\infty$. Hence
we can work in the gauge $A_{z}=0$ for the instanton profile,

\begin{equation}
\left(h^{+},h^{-}\right)=\left(Pe^{-\int_{-\infty}^{+\infty}A_{z}dz},1\right).
\end{equation}
Note that the CS term in (\ref{eq:48}) has to be replaced by the
new CS term in (\ref{eq:50}) where the new contribution can be accordingly
identified as the constraint of the hypercharge (\ref{eq:49}).

Then motion of the collective coordinates can be characterized by
the effective Lagrangian of the collective modes in the moduli space,

\begin{equation}
L=\frac{m_{X}}{2}g_{\alpha\beta}\dot{X}^{\alpha}\dot{X}^{\beta}-U\left(X^{\alpha}\right)+\mathcal{O}\left(\lambda^{-1}\right).\label{eq:53}
\end{equation}
The derivative term in (\ref{eq:53}) refers to the line element in
the moduli space which denotes the kinetics of the collective modes.
Integrating over $\left(x,z\right)$, we obtain the following Hamiltonian
associated to (\ref{eq:53}) which is,

\begin{equation}
H=M_{0}+H_{X}+H_{Z}+H_{\rho}+\mathcal{O}\left(\lambda^{-1}\right),\label{eq:54}
\end{equation}
where,

\begin{align}
H_{X} & =\frac{1}{2m_{X}}P_{X}^{2}+M_{0},\ H_{Z}=\frac{1}{2m_{Z}}P_{Z}^{2}+\frac{1}{2}m_{Z}\omega_{Z}^{2}Z^{2},\nonumber \\
H_{\rho} & =-\frac{1}{2m_{y}}P_{y}^{2}+\frac{1}{2}m_{y}\omega_{y}^{2}\rho^{2}+\frac{\mathcal{Q}}{\rho^{2}}+\frac{2}{m_{y}\rho^{2}}\left(\sum_{s=1}^{3}J_{s}^{2}+2\sum_{s=4}^{7}J_{s}^{2}\right),\label{eq:55}
\end{align}
with the constraint of the hypercharge (\ref{eq:49}) and,

\begin{align}
m_{X} & =m_{z}=\frac{1}{2}m_{y}=8ab^{3/2}\pi^{2}N_{c},\nonumber \\
\omega_{Z} & =\sqrt{\frac{3-b}{3}},\ \omega_{\rho}=\omega_{y}=\sqrt{\frac{3-b}{12}},\nonumber \\
M_{0} & =8ab^{3/2}\pi^{2}\lambda N_{c},\ \mathcal{Q}=\frac{N_{c}}{40ab^{3/2}\pi^{2}},\nonumber \\
\rho^{2} & =\sum_{a=1}^{a=\eta+1}\left(y^{a}\right)^{2}.
\end{align}
Then quantization of the Hamiltonian would be nothing but replacing
the momentum by its derivative operator. Specifically, we need

\begin{equation}
P_{X}^{2}=-\frac{1}{2m_{X}}\sum_{i=1}^{3}\frac{\partial^{2}}{\partial X_{i}^{2}},\ P_{Z}^{2}=-\frac{1}{2m_{Z}}\frac{\partial^{2}}{\partial Z^{2}},\ P_{y}^{2}=-\frac{1}{2m_{y}}\frac{1}{\rho^{\eta}}\partial_{\rho}\left(\rho^{\eta}\partial_{\rho}\right),
\end{equation}
 for the Hamiltonian (\ref{eq:54}) and (\ref{eq:55}).

\subsection{Baryon spectrum}

The baryon spectrum can be obtained by solving the eigen equation
of the Hamiltonian for the collective modes. Let us consider the solution
for the Hamiltonian (\ref{eq:54}) in the $\left(p,q\right)$ representation
for the $SU\left(3\right)_{I}$ and $SU\left(3\right)_{J}$. In this
representation, a state takes the following relation of the quantum
number,

\begin{align}
\sum_{s=1}^{8}\left(J_{s}\right)^{2} & =\frac{1}{3}\left(p^{2}+q^{2}+pq\right)+p+q,\nonumber \\
\sum_{s=1}^{3}\left(J_{s}\right)^{2} & =j\left(j+1\right).
\end{align}
Then the radius part of the Hamiltonian $H_{\rho}$ can be rewritten
as,

\begin{equation}
H_{\rho}=-\frac{1}{2m_{y}}\frac{1}{\rho^{\eta}}\partial_{\rho}\left(\rho^{\eta}\partial_{\rho}\right)+\frac{1}{2}m_{y}\omega_{y}^{2}\rho^{2}+\frac{K}{m_{y}\rho^{2}},\label{eq:59}
\end{equation}
where,

\[
K=\frac{N_{c}^{2}}{15}+\frac{4}{3}\left(p^{2}+q^{2}+pq\right)+4\left(p+q\right)-2j\left(j+1\right),
\]
 and we have used the definition of $J_{8}$ as (\ref{eq:49}). So
the eigenfunction of (\ref{eq:59}) takes the following forms,

\begin{equation}
H_{\rho}\psi\left(\rho\right)=E_{\rho}\psi\left(\rho\right),
\end{equation}
with

\begin{align}
\psi\left(\rho\right) & =e^{-v/2}v^{\mathcal{B}}\gamma\left(v\right),\nonumber \\
v & =m_{y}\omega_{\rho}\rho^{2},\nonumber \\
\mathcal{B} & =\frac{1}{4}\sqrt{\left(\eta-1\right)^{2}+8K}-\frac{1}{4}\left(\eta-1\right).
\end{align}
The function $\gamma\left(v\right)$ needs to be solved by the following
hypergeometrically differential equation,

\begin{equation}
\left[v\frac{d^{2}}{dv^{2}}+\left(2\mathcal{B}+\frac{\eta+1}{2}-v\right)\frac{d}{dv}+\left(\frac{E_{\rho}}{2\omega_{\rho}}-\mathcal{B}-\frac{\eta+1}{4}\right)\right]\gamma\left(v\right)=0.\label{eq:62}
\end{equation}
The normalizable regular solution for (\ref{eq:62}) has to satisfy,

\begin{equation}
\frac{E_{\rho}}{2\omega_{\rho}}-\mathcal{B}-\frac{\eta+1}{4}=n_{\rho}=0,1,2,3...
\end{equation}
so the eigenvalues are given as,

\begin{equation}
E_{\rho}=\omega_{\rho}\left[2n_{\rho}+\frac{\sqrt{\left(\eta-1\right)^{2}+8K}}{2}+1\right].\label{eq:64}
\end{equation}
On the other hand, the $Z$ part Hamiltonian $H_{Z}$ takes the same
form as the Hamiltonian of a harmonic oscillator. Therefore the eigenvalues
of $H_{Z}$ can be immediately obtained as,

\begin{equation}
E_{Z}=\omega_{Z}\left(n_{Z}+\frac{1}{2}\right),\ n_{Z}=0,1,2,3...\label{eq:65}
\end{equation}
Combining (\ref{eq:64}) with (\ref{eq:65}), we finally have the
baryon spectrum as,

\begin{equation}
M=M_{0}+\left(\frac{3-b}{2}\right)^{1/2}\left[\sqrt{\frac{\left(\eta-1\right)^{2}}{24}+\frac{K}{3}}+\sqrt{\frac{2}{3}}\left(n_{Z}+n_{\rho}+1\right)\right]M_{KK}.\label{eq:66}
\end{equation}
Particularly, we put $N_{c}=3$ into the constraint (\ref{eq:49})
which implies,

\begin{equation}
p+2q=3\times\left(\mathrm{integer}\right).\label{eq:67}
\end{equation}
So the low-energy baryon states satisfying the constraint (\ref{eq:49})
and (\ref{eq:67}) takes the following values for the quantum number,

\begin{align}
\left(p,q\right)=\left(1,1\right),\  & j=\frac{1}{2},K=\frac{111}{10},\ \left(\mathrm{octet}\right)\nonumber \\
\left(p,q\right)=\left(3,0\right),\  & j=\frac{3}{2},K=\frac{171}{10},\ \left(\mathrm{decuplet}\right)\nonumber \\
\left(p,q\right)=\left(0,3\right),\  & j=\frac{1}{2},K=\frac{231}{10}.\ \left(\mathrm{anti-decuplet}\right)
\end{align}

\section{Involving the heavy flavor}

\subsection{The heavy flavor brane}

The holographic baryon in the D0-D4/D8 construction has been discussed
in the previous sections, let us extend the analysis to involve the
heavy flavor in this section.

\begin{figure}
\begin{centering}
\includegraphics[scale=0.5]{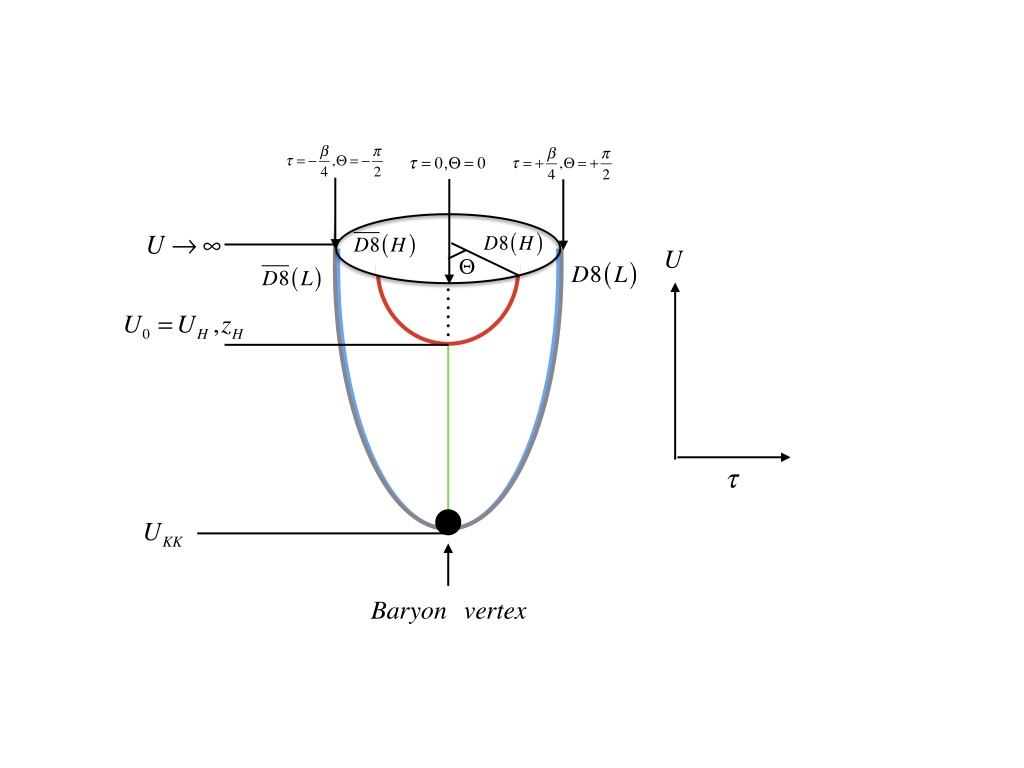}
\par\end{centering}
\caption{\label{fig:2}Various brane configuration in the $\tau-U$ plane where
$\tau$ is compactified on $S^{1}$. The bubble background (cigar)
is produced by $N_{c}$ D4-branes with $N_{0}$ smeared D0-branes.
The $N_{f}=3$ light (chiral) $\mathrm{D}8/\overline{\mathrm{D}8}$-branes
(L) at antipodal position of the cigar are represented by the blue
line. A pair of heavy $\mathrm{D}8/\overline{\mathrm{D}8}$-brane
(H) is separated from the light branes which is represented by the
red line. The massive state is produced by the strings stretched between
the light and heavy branes which is represented by the green line
(i.e. the HL-string) in this figure. The baryon spectrum is produced
by the baryon vertex which lives inside the L-branes.}
\end{figure}

Following \cite{key-015,key-016,key-017,key-018,key-019}, the heavy
flavor could be introduced into this system by embedding a pair of
flavor brane which is separated from the other $N_{f}$ coincident
flavor branes with a string stretched between them as illustrated
in Figure \ref{fig:2}. The $N_{f}$ coincident flavor branes, mentioned
in the previous sections, are now named as ``light flavor branes
(L-brane)'' while the separated flavor brane is named as ``heavy
flavor branes (H-brane)''. So the string stretched between L- and
H-branes is therefore named as ``HL-string'' which produces massive
multiplets. The low-energy modes of the HL-string, which corresponds
to the heavy-light mesons, could be approximated in bi-fundamental
representation by the local vector fields in the vicinity of the light
flavor branes. The heavy-light fields become massive due to the nonzero
vacuum expectation value (vev) of the HL-string. And their mass term
in the action comes from the moduli span by the dilaton fields. For
the heavy flavor branes, we should choose another solution from (\ref{eq:17})
denoted by $\tau_{H}\left(U\right)$ as the embedding function with
$U_{0}=U_{H}\neq U_{KK}$ since they must be embedded at the non-antipodal
position of the background. Accordingly, there is a finite separation
at $\tau_{H}\left(U_{0}\right)=0$ between the H- and L-branes as
shown in Figure \ref{fig:2}. 

\subsection{The effective heavy-light action}

By involving the heavy-light interaction, the subject of this subsection
is to study the effective low-energy dynamics of the baryons on the
L-branes since the baryon vertex lives inside the L-brane. The lowest
modes of the HL-string consist of longitudinal modes $\varPhi_{a}$
and the transverse modes $\varphi^{I}$ in the L-brane world volume.
These fields acquire a nonzero vev at finite brane separation thus
introduce the mass to the vector field \cite{key-020}. These fields
are always mentioned as ``bi-local fields'', however we are going
to approximate them by the local vector fields near the worldvolume
of the L-branes so that their dynamics could be described by the DBI
action. Hence this construction is distinct from the approaches in
\cite{key-021,key-022,key-023,key-024,key-025,key-026,key-027}. 

By keeping these in mind, let us start with the effective action of
the L-branes involving the heavy flavor. The generic expansion of
the DBI for a D8-branes in the leading order can be written as,

\begin{align}
S_{\mathrm{DBI}}^{\mathrm{D8/\overline{D8}}} & =-\frac{T_{8}\left(2\pi\alpha^{\prime}\right)^{2}}{4}\int_{\mathrm{D8/\overline{D8}}}d^{9}\xi\sqrt{-\det G}e^{-\mathbf{\Phi}}\mathrm{Tr}\left\{ \mathcal{F}_{ab}\mathcal{F}^{ab}-2D_{a}\varphi^{I}D_{a}\varphi^{J}+\left[\varphi^{I},\varphi^{J}\right]^{2}\right\} .\nonumber \\
 & \equiv\boldsymbol{S}_{\mathrm{YM}}^{\mathrm{D8/\overline{D8}}}+\boldsymbol{S}_{\Psi}\label{eq:69}
\end{align}
where $\varphi^{I}$ is the transverse mode and the index $a,b$ runs
over the L-brane. Let us define $\varphi^{I}\equiv\Psi$ to omit the
index since there is only one transverse coordinate to D8-brane. The
field $\Psi$ is traceless in adjoint representation additional to
the adjoint gauge field $\mathcal{A}_{a}$. While $\boldsymbol{S}_{\mathrm{YM}}^{\mathrm{D8/\overline{D8}}}$
refers to the Yang-Mills action in (\ref{eq:20}), the worldvolume
field should be combined in a superconnection according to string
theory since the pair of the H-brane is separated from the $N_{f}$
L-branes with a HL-string stretched between them \cite{key-79}. Hence
we use the following matrix-valued 1-form for the gauge field by involving
the heavy flavor, 

\begin{equation}
\boldsymbol{\mathcal{A}}_{a}=\left(\begin{array}{cc}
\mathcal{A}_{a} & \varPhi_{a}\\
-\varPhi_{a}^{\dagger} & 0
\end{array}\right).\label{eq:70}
\end{equation}
Here $\boldsymbol{\mathcal{A}}_{a}$ is $\left(N_{f}+1\right)\times\left(N_{f}+1\right)$
matrix-valued field while $\Psi$ and $\mathcal{A}_{a}$ are $N_{f}\times N_{f}$
valued fields. The $\varPhi_{a}$ multiplet is massless if all the
flavor branes are coincident, otherwise $\varPhi_{a}$ could be massive
field. The gauge field strength corresponding to (\ref{eq:70}) is,

\begin{equation}
\boldsymbol{\mathcal{F}}_{ab}=\left(\begin{array}{cc}
\mathcal{F}_{ab}-\varPhi_{[a}\varPhi_{b]}^{\dagger} & \partial_{[a}\varPhi_{b]}+\mathcal{A}_{[a}\varPhi_{b]}\\
\partial_{[a}\varPhi_{b]}^{\dagger}+\varPhi_{[a}^{\dagger}\mathcal{A}_{b]} & -\varPhi_{[a}^{\dagger}\varPhi_{b]}
\end{array}\right).\label{eq:71}
\end{equation}
Therefore the Yang-Mills action $\boldsymbol{S}_{\mathrm{YM}}^{\mathrm{D8/\overline{D8}}}$
involving the heavy flavor takes the same formula as in (\ref{eq:20})
but replacing the gauge field strength by (\ref{eq:71}), i.e. $\mathcal{F}_{ab}\rightarrow\boldsymbol{\mathcal{F}}_{ab}$.
Using the dimensionless variables defined in (\ref{eq:23}) and rescaling
(\ref{eq:24}), the Yang-Mills Lagrangian associated to action $\boldsymbol{S}_{\mathrm{YM}}^{\mathrm{D8/\overline{D8}}}$
is calculated as,

\begin{equation}
\boldsymbol{\mathcal{L}}_{\mathrm{YM}}^{\mathrm{D8/\overline{D8}}}=\mathcal{L}_{\mathrm{YM}}^{L}+aN_{c}b^{3/2}\lambda\mathcal{L}_{0}^{H}+aN_{c}b^{3/2}\mathcal{L}_{1}^{H}+\mathcal{O}\left(\lambda^{-1}\right),\label{eq:72}
\end{equation}
where $\mathcal{L}_{\mathrm{YM}}^{L}$ refers to the Lagrangian associated
to the action $S_{\mathrm{YM}}^{\mathrm{D8/\overline{D8}}}$ in (\ref{eq:20})
and,

\begin{align}
\mathcal{L}_{0}^{H}= & -\left(D_{M}\varPhi_{N}^{\dagger}-D_{N}\varPhi_{M}^{\dagger}\right)\left(D_{M}\varPhi_{N}-D_{N}\varPhi_{M}\right)+2\varPhi_{M}^{\dagger}\mathcal{F}^{MN}\varPhi_{N},\nonumber \\
\mathcal{L}_{1}^{H}= & 2\left(D_{0}\varPhi_{M}^{\dagger}-D_{M}\varPhi_{0}^{\dagger}\right)\left(D_{0}\varPhi_{M}-D_{M}\varPhi_{0}\right)-2\varPhi_{0}^{\dagger}\mathcal{F}^{0M}\varPhi_{M}-2\varPhi_{0}^{\dagger}\mathcal{F}^{0M}\varPhi_{M}+\widetilde{\mathcal{L}}_{1}^{H},\nonumber \\
\widetilde{\mathcal{L}}_{1}^{H}= & bz^{2}\left(\frac{5}{6}-\frac{1}{2b}\right)\left(D_{i}\varPhi_{j}-D_{j}\varPhi_{i}\right)^{\dagger}\left(D_{i}\varPhi_{j}-D_{j}\varPhi_{i}\right)\nonumber \\
 & -bz^{2}\left(1+\frac{1}{b}\right)\left(D_{i}\varPhi_{z}-D_{z}\varPhi_{i}\right)^{\dagger}\left(D_{i}\varPhi_{z}-D_{z}\varPhi_{i}\right)\nonumber \\
 & -bz^{2}\left(\frac{5}{3}-\frac{1}{b}\right)\varPhi_{i}^{\dagger}\mathcal{F}^{ij}\varPhi_{j}+bz^{2}\left(1+\frac{1}{b}\right)\left(\varPhi_{z}^{\dagger}\mathcal{F}^{zi}\varPhi_{i}+c.c.\right).\label{eq:73}
\end{align}
Note the derivative is defined as $D_{M}\varPhi_{N}=\partial_{M}\varPhi_{N}+\mathcal{A}_{[M}\varPhi_{N]}$.
On the other hand, the action $\boldsymbol{S}_{\Psi}$ in (\ref{eq:69})
is collected as,

\begin{align}
\boldsymbol{S}_{\Psi} & =-\frac{T_{8}\left(2\pi\alpha^{\prime}\right)^{2}}{4}\int d^{9}\xi\sqrt{-\det G}e^{-\boldsymbol{\Phi}}\mathrm{Tr}\left\{ -2D_{a}\varphi^{i}D_{a}\varphi^{i}+\left[\varphi^{i},\varphi^{j}\right]^{2}\right\} \nonumber \\
 & =\tilde{T}_{8}\int d^{4}xdz\sqrt{-\det G}e^{-\boldsymbol{\Phi}}\mathrm{Tr}\left\{ \frac{1}{2}D_{a}\Psi D_{a}\Psi-\frac{1}{4}\left[\Psi,\Psi\right]^{2}\right\} ,\label{eq:74}
\end{align}
with $D_{a}\Psi=\partial_{a}\Psi+i\left[\mathcal{A}_{a},\Psi\right]$.
According to \cite{key-79,key-030}, the extrema of the potential
contribution or $\left[\Psi,\left[\Psi,\Psi\right]\right]=0$ in (\ref{eq:74})
can define the moduli of $\Psi$. So we can choose the moduli solution
of $\Psi$ with a finite vev $v$ as,

\begin{equation}
\Psi=\left(\begin{array}{cc}
-\frac{v}{N_{f}}\boldsymbol{1}_{N_{f}} & 0\\
0 & v
\end{array}\right),\label{eq:75}
\end{equation}
Imposing the solution (\ref{eq:75}) into (\ref{eq:74}), we have 

\begin{equation}
\boldsymbol{S}_{\Psi}=-aN_{c}b^{3/2}\int d^{4}xdz2m_{H}^{2}\varPhi_{M}^{\dagger}\varPhi_{M}+\mathcal{O}\left(\lambda^{-1}\right),\label{eq:76}
\end{equation}
where $m_{H}=\frac{1}{\sqrt{6}}\frac{N_{f}+1}{N_{f}}vb^{1/4}$. Notice
that we have used the dimensionless $v$ by the replacement $v\rightarrow\frac{M_{KK}^{1/2}}{U_{KK}^{1/2}}v$
in (\ref{eq:76}).

Finally, we still need to derive the formula of the CS action involving
the heavy flavor of the L-brane. Similarly, let us replace the gauge
field strength $\mathcal{F}_{ab}\rightarrow\boldsymbol{\mathcal{F}}_{ab}$
in $S_{\mathrm{CS}}^{\mathrm{D}8/\overline{\mathrm{D}8}}$ (\ref{eq:20}),
then we obtain,
\begin{align}
\boldsymbol{\mathcal{L}}_{\mathrm{CS}}^{\mathrm{D}8/\overline{\mathrm{D}8}}= & \mathcal{L}_{\mathrm{CS}}^{L}\left(\mathcal{A}\right)+\mathcal{L}_{\mathrm{CS}}^{H}+\mathcal{O}\left(\lambda^{-1}\right),\nonumber \\
\mathcal{L}_{\mathrm{CS}}^{H}= & -\frac{iN_{c}}{24\pi^{2}}\left(d\varPhi^{\dagger}\mathcal{A}d\varPhi+d\varPhi^{\dagger}d\mathcal{A}\varPhi+\varPhi^{\dagger}d\mathcal{A}d\varPhi\right)\nonumber \\
 & -\frac{iN_{c}}{16\pi^{2}}\left(d\varPhi^{\dagger}\mathcal{A}^{2}\varPhi+\varPhi^{\dagger}\mathcal{A}^{2}d\varPhi+\varPhi^{\dagger}\mathcal{A}d\mathcal{A}\varPhi+\varPhi^{\dagger}d\mathcal{A}\mathcal{A}\varPhi\right)\nonumber \\
 & -\frac{5iN_{c}}{48\pi^{2}}\varPhi^{\dagger}\mathcal{A}^{3}\varPhi+\mathcal{O}\left(\varPhi^{4},\mathcal{A}\right),\label{eq:77}
\end{align}
where $\mathcal{L}_{\mathrm{CS}}^{L}\left(\mathcal{A}\right)$ refers
to the Lagrangian associated to action $S_{\mathrm{CS}}^{\mathrm{D}8/\overline{\mathrm{D}8}}$
in (\ref{eq:20}). So the action for heavy-light interaction can be
collected through (\ref{eq:72}) (\ref{eq:73}) (\ref{eq:76}) (\ref{eq:77}).

\subsection{The double limit}

Since calculating the complete contributions from the heavy meson
field $\varPhi_{M}$ is difficult, we will work in the limit of $\lambda\rightarrow\infty$
followed by $m_{H}\rightarrow\infty$ (i.e. the ``double limit'')
as in the previous works \cite{key-015,key-018,key-019}. The leading
contributions come from the presented $S_{\mathrm{YM}}^{\mathrm{D}8/\overline{\mathrm{D}8}}$
in (\ref{eq:5}) which is of order $\lambda m_{H}^{0}$, while the
heavy-light interaction Lagrangian $\mathcal{L}_{1}^{H}$ in (\ref{eq:73})
and $\mathcal{L}_{\mathrm{CS}}^{H}$ in (\ref{eq:77}) contribute
to the subleading order $\lambda^{0}m_{H}$. The double limit implies
that we assume the heavy meson field $\varPhi_{M}$ is very massive,
or equivalently, the separation of the H- and L-branes is very large
as shown in Figure \ref{fig:2}. the straight pending string accordingly
takes a value at $z=z_{H}$ which satisfies,

\begin{align}
m_{H} & =\frac{1}{\pi l_{s}^{2}}\lim_{z_{H}\rightarrow\infty}\int_{0}^{z_{H}}dz\sqrt{-g_{00}g_{zz}}\nonumber \\
 & \simeq\frac{1}{\pi l_{s}^{2}}U_{KK}^{1/3}z_{H}^{2/3}+\mathcal{O}\left(z_{H}^{0}\right).\label{eq:78}
\end{align}
Let us rewrite (\ref{eq:78}) with the dimensionless variables by
the replacement $m_{H}\rightarrow m_{H}M_{KK}$, $z_{H}\rightarrow z_{H}U_{KK}$
for convenience, then using (\ref{eq:13}) we get,

\begin{equation}
\frac{m_{H}}{\lambda}=\frac{2b}{9\pi}z_{H}^{2/3}.\label{eq:79}
\end{equation}
In the double limit, we follow \cite{key-015,key-018} by redefining
$\varPhi_{M}=\phi_{M}e^{\pm im_{H}x^{0}}$ with ``-'' for particle
and ``+'' for anti-particle. So the derivative term of $\varPhi_{M}$
can be replaced by $D_{0}\varPhi_{M}\rightarrow\left(D_{0}\pm im_{H}\right)\varPhi_{M}$.
Then we collect the terms of order $\lambda^{0}m_{H}$ from the heavy-light
action which are,

\begin{align}
\mathcal{L}_{m_{H}}= & \mathcal{L}_{1,m}+\mathcal{L}_{\mathrm{CS},m},\nonumber \\
\mathcal{L}_{1,m}= & ab^{3/2}N_{c}\left[4im_{H}\phi_{M}^{\dagger}D_{0}\phi_{M}-2im_{H}\left(\phi_{0}^{\dagger}D_{M}\phi_{M}-c.c.\right)\right],\nonumber \\
\mathcal{L}_{\mathrm{CS},m}= & \frac{m_{H}N_{c}}{16\pi^{2}}\epsilon_{MNPQ}\phi_{M}^{\dagger}\mathcal{F}_{NP}\phi_{Q}=\frac{m_{H}N_{c}}{8\pi^{2}}\phi_{M}^{\dagger}\mathcal{F}_{MN}\phi_{N}.\label{eq:80}
\end{align}

\subsection{The equations of motion and the zero mode of vector}

Let us consider the equations of motion of the heavy meson field $\varPhi_{M}$.
The equations read from the Lagrangians in (\ref{eq:72}) (\ref{eq:73})
(\ref{eq:76}) (\ref{eq:77}),

\begin{equation}
D_{M}D_{M}\varPhi_{N}-D_{N}D_{M}\varPhi_{M}+2\mathcal{F}_{NM}\varPhi_{M}+\mathcal{O}\left(\lambda^{-1}\right)=0.\label{eq:81}
\end{equation}
Notice that $\varPhi_{M}$ is independent on $\varPhi_{0}$, thus
the equation of motion for $\varPhi_{0}$ is,

\begin{equation}
D_{M}\left(D_{0}\varPhi_{M}-D_{M}\varPhi_{0}\right)-\mathcal{F}^{0M}\varPhi_{M}-\frac{1}{64\pi^{2}ab^{3/2}}\epsilon_{MNPQ}\mathcal{K}_{MNPQ}+\mathcal{O}\left(\lambda^{-1}\right)=0,\label{eq:82}
\end{equation}
where the 4-form $\mathcal{K}_{MNPQ}$ is given by,

\begin{equation}
\mathcal{K}_{MNPQ}=\partial_{M}\mathcal{A}_{N}\partial_{P}\varPhi_{Q}+\mathcal{A}_{M}\mathcal{A}_{N}\partial_{P}\varPhi_{Q}+\partial_{M}\mathcal{A}_{N}\mathcal{A}_{P}\varPhi_{Q}+\frac{5}{6}\mathcal{A}_{M}\mathcal{A}_{N}\mathcal{A}_{P}\varPhi_{Q}.
\end{equation}
The equation of motion (\ref{eq:82}) suggests a simplification $D_{M}\varPhi_{M}=0$
in (\ref{eq:80}) is considerable which would imply that the transverse
mode $\varPhi_{M}$ is covariant.

In combination with the constraint equation (\ref{eq:82}), we find
the equation of motion (\ref{eq:81}) are equivalent to the vector
equation of zero-mode in the fundamental representation. To show this,
let us recall the self-dual gauge field strength (\ref{eq:30}). By
defining $f_{MN}=\partial_{[M}\phi_{N]}+\mathcal{A}_{[M}\phi_{N]}$,
the Lagrangian $\mathcal{L}_{0}^{H}$ in (\ref{eq:73}) can be rewritten
in the compact form,

\begin{align}
\mathcal{L}_{0}^{H} & =-f_{MN}^{\dagger}f_{MN}+2\phi_{M}^{\dagger}\mathcal{F}_{MN}\phi_{N}\nonumber \\
 & =-f_{MN}^{\dagger}f_{MN}+2\epsilon_{MNPQ}\phi_{M}^{\dagger}D_{N}D_{P}\phi_{Q}\nonumber \\
 & =-f_{MN}^{\dagger}f_{MN}+f_{MN}^{\dagger}\star f_{MN}\nonumber \\
 & =-\frac{1}{2}\left(f_{MN}-\star f_{MN}\right)^{\dagger}\left(f_{MN}-\star f_{MN}\right),\label{eq:84}
\end{align}
where $\star$ denotes the Hodge dual. Hence the equations of motion
(\ref{eq:81}) can be replaced by,

\begin{align}
f_{MN}-\star f_{MN}=0 & ,\nonumber \\
D_{M}\phi_{M}=0 & ,\label{eq:85}
\end{align}
or equivalently,

\begin{equation}
\sigma_{M}D_{M}\psi=0,\ \ \ \ \mathrm{with}\ \ \psi=\bar{\sigma}_{M}\phi_{M}.\label{eq:86}
\end{equation}
Therefore the solution for $\phi_{M}$ in (\ref{eq:85}) can be given
as,

\begin{equation}
\phi_{M}=\bar{\sigma}_{M}\xi\frac{\rho}{\left(x^{2}+\rho^{2}\right)^{3/2}}\equiv\bar{\sigma}_{M}f\left(x\right)\xi,\label{eq:87}
\end{equation}
Here $\xi$ denotes a two-component spinor and the (\ref{eq:87})
agrees with the results in \cite{key-015,key-018}. The interplay
of (\ref{eq:86}) is remarkable because it illustrates that a heavy
vector meson binds to an flavored bulk instanton in holography which
concludes that the zero mode of a vector can be equivalently described
by a spinor.

\subsection{Quantization of the collective modes}

The quantization of the leading $\lambda N_{c}$ contribution has
been discussed in Section 3.2 where the classical moduli of the bound
instanton is quantized by slowly translating and rotating the bound
states. The quantization involving the heavy flavor follows the similar
procedures by replacing $\mathcal{A}\rightarrow\boldsymbol{\mathcal{A}}$.
So we have,

\begin{align}
\varPhi_{M} & \rightarrow V\left[a^{s}\left(t\right)\right]\varPhi_{M}\left[X^{0}\left(t\right),Z\left(t\right),\rho\left(t\right),\chi\left(t\right)\right],\nonumber \\
\varPhi_{0} & =0,\label{eq:88}
\end{align}
where $X^{0}$, $Z$ denotes the center in the $x^{i}$ and $z$ directions
respectively. For the $N_{f}=3$ case, $a^{s}$ represents the $SU\left(3\right)$
gauge rotation. Then the time-dependent configuration could be introduced
in the heavy-light effective action as described earlier.

Due to the additional CS terms in (\ref{eq:50}), it picks up the
collective instanton in quantization by defining,

\begin{align}
h^{-} & =diag\left(a^{s}\left(t\right)^{-1},1\right),\nonumber \\
h^{+} & =h_{0}diag\left(a^{s}\left(t\right)^{-1},1\right).\label{eq:89}
\end{align}
Note that the field $\boldsymbol{\mathcal{A}}$ consists of the instanton
solution $A$ and the zero-mode solution $\Phi$ which carries the
same topological number. Then inserting (\ref{eq:89}) into the new
CS terms (\ref{eq:50}), we obtain,

\begin{equation}
\boldsymbol{S}_{\mathrm{CS}}^{\mathrm{new}}=S_{\mathrm{CS}}-\frac{iN_{c}}{48\pi^{2}}\int_{M_{4}}dt\mathrm{Tr}\left[\mathbf{a}\left(t\right)^{-1}\cdot\dot{\mathbf{a}}\left(t\right)\right]\left(h_{0}^{-1}dh_{0}\right)^{3},
\end{equation}
Here the heavy-light contributions $S_{\mathrm{CS}}$ refers those
in (\ref{eq:77}) while the new contribution from the second term
can be identical to the coupling with the hypercharge as (\ref{eq:49}).

In order to obtain the quantized Lagrangian of the collective modes
in $\lambda^{0}m_{H}$ order, we recall (\ref{eq:80}) by imposing
(\ref{eq:30}) and (\ref{eq:88}) which becomes,

\begin{align}
\mathcal{L}_{m}= & ab^{3/2}N_{c}\bigg[16im_{H}\xi^{\dagger}\partial_{t}\xi f^{2}-16m_{H}\xi^{\dagger}\xi f^{2}\frac{2\sqrt{6}+1}{6}A_{0}\nonumber \\
 & -m_{H}f^{2}\xi^{\dagger}\sigma_{\mu}\Phi\bar{\sigma}_{\mu}\xi+m_{H}\xi^{\dagger}\xi f^{2}\frac{3}{ab^{3/2}\pi^{2}}\frac{\rho^{2}}{\left(x^{2}+\rho^{2}\right)^{2}}\bigg].\label{eq:91}
\end{align}
The second term comes from the $A_{0}$ coupling and it would simply
the expression for the zero-modes by,

\begin{equation}
\xi^{\dagger}\sigma_{\mu}\Phi\bar{\sigma}_{\mu}\xi=a_{8}\frac{8\xi^{\dagger}\xi}{\sqrt{3}}.
\end{equation}
The last term in (\ref{eq:91}) originates from the CS term in (\ref{eq:80})
with,

\begin{equation}
\frac{im_{H}N_{c}}{8\pi^{2}}\phi_{M}^{\dagger}F_{MN}\phi_{N}=\frac{3im_{H}N_{c}}{\pi^{2}}\frac{f^{2}\rho^{2}}{\left(x^{2}+\rho^{2}\right)^{2}}\xi^{\dagger}\xi.\label{eq:93}
\end{equation}
There are additional terms to (\ref{eq:91}) due to the $\xi^{\dagger}\xi$
coupling to the $U(1)$ gauge field $A_{0}$ which induces a Coulomb-like
backreaction of the form $\left(\xi^{\dagger}\xi\right)^{2}$ as we
have indicated in \cite{key-015}. By keeping this in mind and using
the normalization $\int d^{4}xf^{2}=1$, the explicit form of $\mathcal{A}_{0}^{\mathrm{cl}}$
and the rescaling $\xi\rightarrow\xi/\sqrt{16ab^{3/2}N_{c}m_{H}}$,
we finally obtain,

\begin{align}
\mathcal{L}= & \mathcal{L}^{L}\left[a_{I},X^{\alpha}\right]+i\xi^{\dagger}\partial_{t}\xi+\frac{\sqrt{6}+2}{48\pi^{2}ab^{3/2}\rho^{2}}\xi^{\dagger}\xi\nonumber \\
 & -\frac{13\left(\xi^{\dagger}\xi\right)^{2}}{288\pi^{2}ab^{3/2}\rho^{2}N_{c}}+a^{8}\frac{N_{c}}{2\sqrt{3}}\left(1-\frac{\xi^{\dagger}\xi}{N_{c}}\right),\label{eq:94}
\end{align}
where $\mathcal{L}^{L}\left[a_{I},X^{\alpha}\right]$ refers to the
Lagrangian (\ref{eq:53}) of light flavors. The (\ref{eq:94}) can
be understood as Lagrangian density of heavy-light degrees of freedom
supplemented by the constraint of hypercharge with the excitations
of the vacuum, namely,

\begin{equation}
\mathcal{L}=\mathcal{L}^{L}\left[a_{I},X^{\alpha}\right]+i\xi^{\dagger}\partial_{t}\xi+\frac{\sqrt{6}+2}{48\pi^{2}ab^{3/2}\rho^{2}}\xi^{\dagger}\xi-\frac{13\left(\xi^{\dagger}\xi\right)^{2}}{288\pi^{2}ab^{3/2}\rho^{2}N_{c}},\label{eq:95}
\end{equation}
with the hypercharge constraint,

\begin{equation}
J^{8}=\frac{N_{c}}{2\sqrt{3}}\left(1-\frac{\xi^{\dagger}\xi}{N_{c}}\right),
\end{equation}

\subsection{The heavy-light baryon spectrum}

The heavy-light Hamiltonian associated to (\ref{eq:95}) takes the
following form,

\begin{equation}
\mathcal{H}=\mathcal{H}^{L}\left[a_{I},X^{\alpha}\right]-\frac{\sqrt{6}+2}{48\pi^{2}ab^{3/2}\rho^{2}}\xi^{\dagger}\xi+\frac{13\left(\xi^{\dagger}\xi\right)^{2}}{288\pi^{2}ab^{3/2}\rho^{2}N_{c}},\label{eq:97}
\end{equation}
with the quantization rule for the spinor $\xi$,

\begin{equation}
\xi_{i}\xi_{j}^{\dagger}+\xi_{j}^{\dagger}\xi_{i}=\delta_{ij}.
\end{equation}
Here $\mathcal{H}^{L}\left[a_{I},X^{\alpha}\right]$ refers to the
Hamiltonian (\ref{eq:54}). Let us use $U$ and $\Lambda$ to represent
the rotation of a spinor and a vector respectively i.e. $\xi\rightarrow U\xi,\ \phi_{M}\rightarrow\Lambda_{MN}\phi_{N}$,
so we have the transformation $U^{-1}\bar{\sigma}_{M}U=\Lambda_{MN}\bar{\sigma}_{M}$
which implies the rotation of the spinor $\xi$ is equivalent to a
spatial rotation of the heavy vector meson field $\phi_{M}$. Notice
that the parity of $\xi$ is negative which is opposite to $\phi_{M}$. 

The spectrum of (\ref{eq:97}) follows the same discussion in Section
3. Since the Hamiltonian (\ref{eq:54}) contains two terms which are
proportional to $\rho^{-2}$, the heavy-light spectrum can be obtained
by modifying $Q$ if comparing (\ref{eq:97}) with Hamiltonian (\ref{eq:54}),
which is,

\begin{equation}
Q=\frac{N_{c}}{40ab^{3/2}\pi^{2}}\rightarrow\frac{N_{c}}{40ab^{3/2}\pi^{2}}\left[1-\frac{5\sqrt{6}+10}{6N_{c}}\xi^{\dagger}\xi+\frac{65}{36N_{c}^{2}}\left(\xi^{\dagger}\xi\right)^{2}\right].\label{eq:99}
\end{equation}
Let us use $\boldsymbol{\mathrm{J}}$ and $\boldsymbol{\mathrm{I}}$
to denote the total spin and isospin, the relation of them is given
by,

\begin{equation}
\vec{\boldsymbol{\mathrm{J}}}=-\vec{\boldsymbol{\mathrm{I}}}_{SU\left(2\right)}+\vec{\boldsymbol{\mathrm{S}}}_{\xi}=-\vec{\boldsymbol{\mathrm{I}}}_{SU\left(2\right)}+\xi^{\dagger}\frac{\vec{\tau}}{2}\xi.
\end{equation}
Here $\vec{\boldsymbol{\mathrm{I}}}_{SU\left(2\right)}$ refers to
the first three generators in the induced representation for a general
$SU(3)$ group. The quantum states for a single bound state i.e. $N_{Q}\equiv\xi^{\dagger}\xi=1$
and $IJ^{\pi}$ assignments are labeled by,

\begin{equation}
|N_{Q},p,q,j,n_{Z},n_{\rho}>\ \ with\ \ IJ^{\pi}=\frac{l}{2}\left(\frac{l}{2}\pm\frac{1}{2}\right)^{\pi}.
\end{equation}
Here $n_{Z},n_{\rho}=0,1,2...$ respectively denotes the number of
quanta associated to the collective motion and the radial breathing
of the instanton core. According to (\ref{eq:66}) and (\ref{eq:99}),
the spectrum of the bound heavy-light state in the D0-D4/D8 system
is given as,

\begin{equation}
M_{N_{Q}}=M_{0}+N_{Q}m_{H}+\left(\frac{3-b}{2}\right)^{1/2}\left[\sqrt{\frac{2}{3}}\left(n_{\rho}+n_{z}+1\right)+\sqrt{\frac{49}{24}+\frac{\boldsymbol{\mathrm{K}}}{3}}\right]M_{KK},\label{eq:102}
\end{equation}
where,

\begin{align}
\boldsymbol{\mathrm{K}}= & \frac{2N_{c}^{2}}{5}\left(1-\frac{5\sqrt{6}+10}{6}\frac{N_{Q}}{N_{c}}+\frac{65}{36}\frac{N_{Q}^{2}}{N_{c}^{2}}\right)-\frac{N_{c}^{2}}{3}\left(1-\frac{N_{Q}}{N_{c}}\right)^{2}\nonumber \\
 & +\frac{4}{3}\left(p^{2}+q^{2}+pq\right)+4\left(p+q\right)-2j\left(j+1\right),
\end{align}
and $M_{0}=\frac{\lambda N_{c}b^{3/2}}{27\pi}M_{KK}$, $M_{KK}$ is
the Kaluza-Klein mass. 

The heavy baryon also includes anti-heavy quarks. So in order to amount
an anti-heavy-light meson, we return to the preceding arguments of
the reduction $\varPhi_{M}=\phi_{M}e^{+im_{H}x^{0}}$. Most of the
calculations would be unchanged as we have indicated in the previous
articles except for pertinent minus signs to the effective Lagrangian.
If we bind one heavy-light and one anti-heavy-light meson in the form
of a zero-mode, the effective Lagrangian now reads,

\begin{align}
\mathcal{L}= & \mathcal{L}^{L}\left[a_{I},X^{\alpha}\right]+i\xi_{Q}^{\dagger}\partial_{t}\xi_{Q}+\frac{\sqrt{6}+2}{48\pi^{2}ab^{3/2}\rho^{2}}\xi_{Q}^{\dagger}\xi_{Q}\nonumber \\
 & -i\xi_{\bar{Q}}^{\dagger}\partial_{t}\xi_{\bar{Q}}-\frac{\sqrt{6}+2}{48\pi^{2}ab^{3/2}\rho^{2}}\xi_{\bar{Q}}^{\dagger}\xi_{\bar{Q}}+\frac{13\left(\xi_{Q}^{\dagger}\xi_{Q}-\xi_{\bar{Q}}^{\dagger}\xi_{\bar{Q}}\right)^{2}}{288\pi^{2}ab^{3/2}\rho^{2}N_{c}}.
\end{align}
Therefore the mass spectrum of baryons with $N_{Q}$ heavy-quarks
and $N_{\bar{Q}}$ anti-heavy quarks can be obtained by following
the similar analysis with the substitution $N_{Q}\rightarrow N_{Q}-N_{\bar{Q}}$
in the order of $\lambda^{0}m_{H}$, so its explicit form is,

\begin{equation}
M_{N_{Q}}=M_{0}+\left(N_{Q}-N_{\bar{Q}}\right)m_{H}+\left(\frac{3-b}{2}\right)^{1/2}\left[\sqrt{\frac{2}{3}}\left(n_{\rho}+n_{z}+1\right)+\sqrt{\frac{49}{24}+\frac{\boldsymbol{\mathrm{K}}}{3}}\right]M_{KK},
\end{equation}
where

\begin{align}
\boldsymbol{\mathrm{K}}= & \frac{2N_{c}^{2}}{5}\left[1-\frac{5\sqrt{6}+10}{6}\frac{N_{Q}-N_{\bar{Q}}}{N_{c}}+\frac{65}{36}\frac{\left(N_{Q}-N_{\bar{Q}}\right)^{2}}{N_{c}^{2}}\right]-\frac{N_{c}^{2}}{3}\left(1-\frac{N_{Q}-N_{\bar{Q}}}{N_{c}}\right)^{2}\nonumber \\
 & +\frac{4}{3}\left(p^{2}+q^{2}+pq\right)+4\left(p+q\right)-2j\left(j+1\right).
\end{align}

\section{Summary and discussion}

In this paper, we have investigated the baryon spectrum in the Witten-Sakai-Sugimoto
model with the D0-D4 background in the situation of three flavors.
The baryon states are created by the baryon vertex which could be
treated as the instanton configuration of the gauge fields on the
flavor branes, so we particularly study the quantization of the collective
modes with the instanton solution. Then the baryon spectrum can be
obtained by solving the Hamiltonian of the collective modes. One of
the key importance here is that the original CS term has to be modified
in order to reproduce the hypercharge constraint as discussed in \cite{key-028,key-029}.
Afterwards we extend our analysis to include the heavy flavor. By
employing the mechanism proposed in \cite{key-017,key-018,key-019},
the heavy flavor is introduced by embedding an extra pair of flavor
branes (heavy flavor brane) which is separated from the other coincident
flavor branes (light flavor branes) with a string stretched between
them. We derive the effective dynamics of the baryons involving the
heavy flavor in the double limits and find that heavy meson binds
in the zero mode of the flavor instanton. The quantization in the
presence of heavy flavors follows the similar procedures as discussed
in the sector of light flavors and we finally obtain the baryon spectra
for the single- and double-heavy baryon all in the case of $N_{f}=3$. 

Since the smeared D0-branes are D-instantons in the background geometry,
this model is holographically dual to the confined Yang-Mills theory
with an excitation of nonzero glue condensate $\left\langle \mathrm{Tr}\mathcal{F}\wedge\mathcal{F}\right\rangle =8\pi^{2}N_{c}\tilde{\kappa}$.
Therefore the baryon spectrum depends on such excitations through
a parameter $b$ and it allows to describe the influence of some metastable
states with odd $P$ or $CP$ parity which may probably be produced
in the hot and dense situation in RHIC. We notice that the constraint
of the parameter $b$ has to satisfy $1\leq b<3$ and the difference
in the spectra becomes larger as the true vacuum appears in the dual
field theory. Specifically, our results would return to those in \cite{key-028,key-019}
if $b=1$ i.e. no D0-branes, and the spectrum would become complex
if $b>3$. Accordingly, it concludes that baryon can not stably exist
if the glue condensate is sufficiently large, which is in agreement
with the previous studies of this model \cite{key-07,key-011,key-012,key-015}.

Last but not the least, let us briefly discuss the comparison with
experimental data by using our holographic baryon spectra. Notice
that we will set $N_{c}=3,\eta=8$ in our spectrum to fit three flavor
QCD. In the sector of light flavors, the spectrum (\ref{eq:66}) shows
the following mass difference between the octet and the decuplet baryon
states, and that between the octet and the anti-decuplet states all
with the same ($n_{\rho},n_{Z}$):

\begin{align}
M_{\mathbf{10}}-M_{\mathbf{8}} & \simeq0.27309\sqrt{3-b}M_{KK},\nonumber \\
M_{\mathbf{10}^{*}}-M_{\mathbf{8}} & \simeq0.51264\sqrt{3-b}M_{KK},
\end{align}
which indicates the metastable states of baryon or the dependence
of the D0 charge. To realize the experimental data,

\begin{align}
M_{\mathbf{10}}^{\mathrm{exp}}-M_{\mathbf{8}}^{\mathrm{exp}} & \simeq292\mathrm{MeV},\nonumber \\
M_{\mathbf{10}^{*}}^{\mathrm{exp}}-M_{\mathbf{8}}^{\mathrm{exp}} & \simeq590\mathrm{MeV},
\end{align}
with taking account into the value of $M_{KK}$ which is determined
from $\rho$ meson mass $M_{KK}=949\mathrm{MeV}$ \cite{key-05},
we can fit the experimental data by setting $b=1.7305$. However $b$
in this method has been identified as an arbitrarily phenomenal parameter.
Strictly speaking, the stable baryon state corresponds to $b=1$.
So in this sense, the above baryon spectrum gives,

\begin{align}
M_{\mathbf{10}}-M_{\mathbf{8}} & \simeq367\mathrm{MeV},\nonumber \\
M_{\mathbf{10}^{*}}-M_{\mathbf{8}} & \simeq688\mathrm{MeV},\label{eq:109}
\end{align}
which is not very close to the experimental data. Since all the quarks
are massless in the light sector, the present three-flavor holographic
baryon spectrum (\ref{eq:66}) might not be very realistic which therefore
induces that (\ref{eq:109}) is a little far away from the experimental
data.

Nonetheless, let us continue the analysis in the presence of the heavy
flavor and find a way to fit the experiments since the mass term has
been introduced in the heavy-light sector. The lowest heavy baryon
states with one heavy quark are characterized by $n_{\rho},n_{Z}=0,1$,
$\left(p,q,j\right)=\left(0,1,0\right)$ , $N_{Q}=1$ for $\bar{\mathbf{3}}$
representation and $\left(p,q,j\right)=\left(2,0,1\right)$ for $\mathbf{6}$
representation. The parity and spin of $\bar{\mathbf{3}}$ states
is $\frac{1}{2}^{+}$, so we can identify these states as $\Sigma_{Q}^{*},\Xi_{Q}\left(\mathbf{6}\right)^{*},\Omega_{Q}^{*}$
respectively. Then the mass spectra are give by (\ref{eq:102}) as,

\begin{align}
M_{\bar{\mathbf{3}}}\left(b\right) & \simeq M_{0}+m_{H}+\left(1.29+\frac{n_{\rho}+n_{Z}+1}{\sqrt{3}}\right)\sqrt{3-b}M_{KK},\nonumber \\
M_{\mathbf{6}}\left(b\right) & \simeq M_{0}+m_{H}+\left(1.53+\frac{n_{\rho}+n_{Z}+1}{\sqrt{3}}\right)\sqrt{3-b}M_{KK},
\end{align}
or equivalently,

\begin{align}
M_{\bar{\mathbf{3}}}\left(b\right)-M\left(b\right)_{p=q=1,N_{Q}=0,j=1/2}-m_{H} & \simeq-0.403\sqrt{3-b}M_{KK},\nonumber \\
M_{\mathbf{6}}\left(b\right)-M\left(b\right)_{p=q=1,N_{Q}=0,j=1/2}-m_{H} & \simeq-0.167\sqrt{3-b}M_{KK},
\end{align}
which implies the mass splitting $M_{\mathbf{6}}\left(b\right)-M_{\bar{\mathbf{3}}}\left(b\right)\simeq0.237\sqrt{3-b}M_{KK}$
become large as those metastable states decay to the true vacuum in
the dual field theory. Since there are also heavy baryon states with
two heavy quarks, we can further evaluate the masses of such states
by following the same construct with $N_{Q}=2$ and $J^{8}=\frac{1}{2\sqrt{3}}$
in (\ref{eq:102}). So the lowest states of heavy baryon binding double
heavy mesons are now characterized by $n_{\rho},n_{Z}=0,1$, $\left(p,q,j\right)=\left(1,0,0\right)$
for the $\mathbf{3}$-plet representation which can be identified
as $\Xi_{QQ}$ with u, d light quark, and $\Omega_{QQ}$ with s quark.
Afterwards, their masses are given as,

\begin{equation}
M_{QQ}^{\mathbf{3}}\left(b\right)-M\left(b\right)_{p=q=1,N_{Q}=0,j=1/2}-2m_{H}\simeq-0.489\sqrt{3-b}M_{KK}.
\end{equation}
For the baryon states binding $N_{Q}$ heavy quarks ($Q$) and $N_{\bar{Q}}$
anti-heavy quarks ($\bar{Q}$), the dependence of the D0 charge could
be obtained by following the same analysis above. For example, the
lowest states are characterized by $N_{Q}=N_{\bar{Q}}=1$, $n_{\rho},n_{Z}=0,1$,
$\left(p,q,j\right)=\left(1,1,\frac{1}{2}\right)$ for the $\mathbf{8}$-plet
representation with $J^{\pi}$ assignments $\frac{1}{2}^{-},\frac{3}{2}^{-}$
and $\left(p,q,j\right)=\left(3,0,\frac{3}{2}\right)$ for $\mathbf{10}$-plet
representation $\frac{5}{2}^{-},\frac{3}{2}^{-},\frac{1}{2}^{-}$.
So the the mass splitting is given as,

\begin{equation}
M_{Q\bar{Q}}^{\mathbf{10}}\left(b\right)-M_{Q\bar{Q}}^{\mathbf{8}}\left(b\right)=0.273\sqrt{3-b}M_{KK}.
\end{equation}

Finally, we expect this model could capture the qualitative behavior
of $\tilde{\kappa}$ (or $\theta$ angle) in QCD-like or Yang-Mills
theory when the heavy-light interaction is involved since the baryon
spectra demonstrate the behavior of baryons as discussed in two-flavor
case \cite{key-01,key-011}. While this model theoretically describes
the influence of the glue condensate in the baryon states, it has
to be further confirmed with more experimental data.

\section*{Acknowledgement}

This work is inspired by \cite{key-07,key-015}. We would like to
acknowledge the support of Fudan University and Shanghai University.
SWL is supported partially by the Thousand Young Talents Program of
China. WHC is supported partially by NSFC China (Grant No. 11375110)
and Shanghai postdoctoral daily foundation.

\end{document}